\documentclass[]{article}

\usepackage{authblk}
\usepackage[T1]{fontenc}
\usepackage[utf8]{inputenc}
\usepackage{amsmath}
\usepackage{amssymb}
\usepackage{amsfonts}
\usepackage{graphicx}
\usepackage{geometry}
\usepackage{multirow}
\usepackage{algorithm}% http://ctan.org/pkg/algorithms
\usepackage{algpseudocode}% http://ctan.org/pkg/algorithmicx
\usepackage{array}
\usepackage{fixltx2e}
\usepackage{url}
\usepackage{color}
\usepackage{xcolor}
\usepackage{scrextend}
\usepackage{tikz}
\usepackage{pdflscape}
\newcolumntype{?}{!{\vrule width 1.3pt}}

\newcommand{\mbf}[1]{\mathbf{#1}}

\title{Sample size calculations for the experimental comparison of multiple algorithms on multiple problem instances\footnote{Submitted for publication in the Journal of Heuristics on August 05, 2019.}}
\author[1,2]{Felipe Campelo}
\author[1]{Elizabeth F. Wanner}

\affil[1]{Aston University, Birmingham B4 7ET, UK. \{f.campelo,e.wanner\}@aston.ac.uk}
\affil[2]{Department of Electrical Engineering, 
  Universidade Federal de Minas Gerais, Brazil.}
  
\begin{document}
\date{August 05, 2019}
\maketitle

\begin{abstract}
This work presents a statistically principled method for estimating the required number of instances in the experimental comparison of multiple algorithms on a given problem class of interest. This approach generalises earlier results by allowing researchers to design experiments based on the desired best, worst, mean or median-case statistical power to detect differences between algorithms larger than a certain threshold. Holm's step-down procedure is used to maintain the overall significance level controlled at desired levels, without resulting in overly conservative experiments. This paper also presents an approach for sampling each algorithm on each instance, based on optimal sample size ratios that minimise the total required number of runs subject to a desired accuracy in the estimation of paired differences. A case study investigating the effect of 21 variants of a custom-tailored Simulated Annealing for a class of scheduling problems is used to illustrate the application of the proposed methods for sample size calculations in the experimental comparison of algorithms.
\end{abstract}

% ===========================================================================
\section{Introduction}
\label{sec:Introduction}
Experimental comparison of algorithms has long been recognised as an essential aspect of research on meta-heuristics  \cite{Barr1995,Hooker1996,McGeoch1996,CoffinSaltzman2000}. Despite well-earned criticisms of some disconnect between theory and experimentation \cite{Hooker1994,Chimani2010} and a few methodological issues that still require adequate attention \cite{Eiben2002,BartzBeielstein2015}, experimental evaluation remains a central aspect and a major component of the development and understanding of EAs. 

This reliance on experimental assessment and comparison of algorithms is evidenced by the continuing effort of researchers in devising better experimental protocols for performance assessment and comparison of algorithms. While many of the most important points were presented as far back as the late 1990s \cite{Barr1995,McGeoch1996,Hooker1996}, research into adequate protocols and tools for comparing algorithms has continued in the past two decades, with several statistical approaches being proposed and employed for comparing the performance of algorithms \cite{CoffinSaltzman2000,Johnson2002,Yuan2004,Demsar2006,YuanGallagher2009,Birattari2004,Birattari2007,Bartz-Beielstein2006,Bartz-Beielstein2010,Garcia2008,Garcia2010,Derrac2011,Carrano2011,Derrac2014,Benavoli2014,Krohling2015,COCO2016,Campelo2019,Calvo2019}. This increased prevalence of more statistically sound experiments in the field of optimisation heuristics can be seen as part of the transition of the area into what has been called the \textit{scientific} period of research on metaheuristics \cite{Sorensen2018}.

Despite of this area-wide effort, a few important aspects of experimental comparisons of algorithms remain largely unexplored in the literature. One of those topics is the question of how to adequately determine the relevant sample sizes for the comparisons of algorithms - number of problem instances to use, and number of runs to employ for each algorithm on each instance. The standard approach has been that of maximizing the number of instances, limited only by the computational budget available \cite{Barr1995,Garcia2008,Garcia2009,Derrac2011,Amo2012}, and of running arbitrarily-set numbers of repeated runs, usually 30 or 50. While it is indeed true that the sensitivity of comparisons increases with the number of instances, this does not mean  that a large sample size can substitute a well-designed experiment \cite{Lenth2001,Mathews2010,Campelo2019}. Also, it is important to be aware that arbitrarily large sample sizes allow tests to detect even minuscule differences at arbitrarily strict significance levels, which may lead to the wrongful interpretation that effects of no practical consequence are strongly significant \cite{Mathews2010,BartzBeielstein2005} if certain methodological safeguards are not put into place when designing the experiment \cite{Campelo2019}. 

The dual issues of statistical power and sample size have been touched by a few authors in the past, albeit only superficially \cite{Jain1991,CoffinSaltzman2000,Czarn2004,Ridge2007,BartzBeielstein2005,Bartz-Beielstein2006,Bartz-Beielstein2010}. Until recently, the best works on the subject were those by Mauro Birattari \cite{Birattari2004,Birattari2009} and Chiarandini and Goegebeur \cite[Ch. 10]{Bartz-Beielstein2010}. The former correctly advocated for a greater focus on the number of instances than on repetitions, showing that the optimal allocation of computational effort, in terms of accuracy in the estimation of mean performance for a given problem class, is to maximize the amount of test instances, running each algorithm a single time on each instance if needed. Birattari's approach, however, was focussed only on accuracy of parameter estimation for the mean performance across a problem class, yielding very little information on the specific behaviour of each  algorithm on each instance, as well as not considering questions of desired statistical power or sample size calculations. The work by Chiarandini and Goegebeur \cite[Ch. 10]{Bartz-Beielstein2010} provided a good discussion on statistical power and sample size in the context of nested linear statistical models, as well as some guidelines on the choice of the number of instances and number of repeated runs based on the graphical analyses of power curves. It was, however, limited to nested models, and required manual inspection of power curves by the user, which may have precluded its broader adoption in the literature.

More recently, we have proposed a principled approach for calculating both the number of instances and number of repeated runs when comparing the performance of two algorithms for a problem class of interest \cite{Campelo2019}. That approach calculates the number of instances based on the desired sensitivity to detect some \textit{minimally relevant effect size} (MRES), i.e., the smallest difference between algorithms that is considered to have some degree of practical relevance. Alternatively, the number of instances can be fixed a priori (e.g., when using standard benchmark sets) and the sensitivity curves can be derived instead. As for the number of repeated runs per instance, the approach proposed in \cite{Campelo2019} was based on minimising the number of algorithm runs required to achieved a the desired accuracy in the estimation of the paired differences in performance for each instance. While statistically sound, the fact that the results presented in that work applied only to comparisons of two algorithms limited their applicability to the general case of experimental comparisons of algorithms, which often aim to investigate the relative performances of multiple algorithms, or multiple variants of a given algorithmic framework.

In this paper we generalise the results from \cite{Campelo2019} to calculate the required sample sizes for the comparison of an arbitrary number of algorithms. The calculation of the number of instances is based on considerations regarding the test of multiple hypotheses that commonly occurs in these contexts, correcting the significance level of each test so as to maintain the familywise error rate controlled at a given desired level \cite{Shaffer1995}. Optimality-based ratios of sample sizes are derived for sampling algorithms on each individual instance, both for the simple difference of means and for two types of percent differences. A simple sampling heuristic is also provided, which can easily generalise the sampling approach presented here for cases in which general performance statistics are of interest.\footnote{All methods presented in this paper are implemented in the form of \texttt{R} package \texttt{CAISEr}, available at\\\url{https://cran.r-project.org/package=CAISEr} \cite{R}.}

The remainder of this paper is organised as follows: Section \ref{sec:probdef} explicitly defines the algorithm comparison problem considered in this work, states the hypotheses to be tested, and provides the rationale for the choices that justify the proposed approach for sample size calculation. Section \ref{sec:nruns} describes the proposed approach for sampling an arbitrary number of algorithms on any given problem instance, and presents the derivation of optimal sample size ratios for three common cases. Section \ref{sec:ninstances} describes the general concepts behind the estimation of the required number of instances for an experiment to achieve desired statistical properties. Section \ref{sec:application} presents an example of application of the proposed methodology for the evaluation of 21 algorithmic variants of a state-of-the-art, custom-tailored simulated annealing approach \cite{Santos2016} for a class of scheduling problems \cite{Vallada2011}. Finally, Section \ref{sec:conclusions} presents final considerations and conclusions.

\section{Problem Definition}
\label{sec:probdef}
% The algorithm comparison problem, multi-algo version
Before proceeding to investigate the required sample sizes for the experimental comparison of algorithms, it is important to formally define the questions one is trying to answer when those comparisons are performed. While there are several different scientific questions that can be investigated by experimentation, arguably the most general case for experimental meta-heuristic comparisons is: \textit{given a finite subset of problem instances and a finite number of runs that can be performed by each algorithm being tested, what can be said regarding the relative performance of those algorithms (according to a given set of performance indicators) for the problem class from which the test instances were drawn?} Notice that this question is broad enough to encompass the most common cases in the experimental comparison of algorithms, and even several comparisons of scientific relevance that are not routinely performed \cite{Eiben2002}. 

Aiming to highlight the specific aspects that are considered in the present work, we refine this problem definition as follows: Let $\Gamma_S = \left\{\gamma_\ell: \ell\in\left[1,N\right]\right\} \subset \Gamma$ represent a finite sample of problem instances drawn from a given problem class of interest, $\Gamma$; and let $\mathcal{A} =\left\{a_1, a_2, \dotsc, a_{A}\right\}$ denote a set of algorithms%
\footnote{Each representing a complete instantiation of a given algorithmic framework, i.e., with both structure and parameter values fully specified} %
that we want to compare. We define the algorithm comparison problem in the context of this work as that of obtaining a (partial) ordering of the algorithms according to their mean performance on the problem class of interest, based on the observed performance values across the set of test instances used. We assume here that (i) all algorithms can be run on the same sample of instances; (ii) any run of an algorithm returns some tentative solution, which can be used to estimate the performance of that algorithm for the problem instance; and (iii) the researcher is interested primarily in comparing the performance of the algorithms for the \textit{problem class} $\Gamma$, instead of for individual instances. Each of these three experimental assumptions can be associated with a specific aspect when comparing algorithms:
\begin{itemize}
	\item Assumption (i) indicates that the variability due to instance effects can be modelled out of our analysis by \textit{pairing} or \textit{blocking} \cite{Campelo2019,Montgomery2013}, which is the underlying approach of methods such as ANOVA with blocking, or Friedman's test \cite{Demsar2006}. 
	\item Assumption (ii) essentially prevents us from having to deal with the problem of missing values in our analysis, since every run will return some valid performance observation. The question on how to deal with missing values in the comparative analysis of algorithms - e.g., if a given algorithm fails to converge in a study where performance is measured by time-until-convergence - is a very relevant one, but it falls outside the scope of this particular work.
	\item Assumption (iii) is associated with the fact that comparative testing of algorithms should, in the vast majority of cases, be focused on generalising the performance observed on limited test set $\Gamma_S$ to the wider class of problems $\Gamma$. This contrasts with a somewhat common practice in the heuristics literature of performing individual Rank Sum tests on each instance and tallying up the wins/losses/ties, without any further inference being performed on these summary quantities. The problem with this approach for the comparison of algorithms on multiple problem instances has been recognised at least since 2006 \cite{Demsar2006}, but despite of this the practice still persists in the literature.
\end{itemize}

Focusing on testing hypotheses regarding the (relative) performance of algorithms for a problem class of interest also determines two other aspects of the design and analysis of comparative experiments. The first is the issue of what is considered the \textit{effective} sample size for these experiments. As  argued in earlier works \cite{Campelo2019,Birattari2004,Bartz-Beielstein2010}, the effective sample size to be considered when testing the hypotheses of interest is the number of \textit{instances}, rather than the amount of repeated runs (or even worse, the product of these two quantities). Failure to account for this when performing inference results in pseudoreplication \cite{Hurlbert1984,Millar2004,Lazic2010}, that is, the (often implicit) use of artificially inflated degrees-of-freedom in statistical tests. This results in violation of the independence assumption underlying these tests, and results in actual confidence levels that can be arbitrarily inferior to the nominal ones. 

The second aspect, which is a consequence of the first, is that the analysis of data generated by the experiments can be done using either a hierarchical model \cite{Gelman2006} or, alternatively, a block design applied on summarised observations, in which the performance of each algorithm on each instance is summarised as a single value, often the mean or median of multiple runs. This second approach is equivalent to the first under the assumption (iii) stated above, and results in the application of the well-known and widely-used methods for comparisons of the average performance of multiple algorithms on multiple problem instances: omnibus tests such as Complete Block Design (CBD) ANOVA or Friedman's test \cite{Demsar2006}. Those tests are used to investigate whether at least one algorithm has an average performance different from at least one other. If that is the case, then multiple comparison procedures are employed to answer the more specific statistical questions of which algorithms differ from which in terms of their average performance \cite{Sheskin2011,Montgomery2013}.

In the following, we detail the hypotheses of interest that are tested by these statistical procedures when performing comparative experiments with algorithms, and suggest that a greater focus on the multiple comparison procedures can provide experimenters with better tools for designing and analysing their experiments.

\subsection{Test Hypotheses}
\label{subsec:TH}

Let $Y_{k\mid\ell}$ denote the performance of algorithm $a_k$ on instance $\gamma_\ell$. Based on the definitions from the preceding section, we can decompose the performance according to the usual model of the complete block design \cite{Montgomery2013b}:
\begin{equation}
\label{eq:statmodel}
Y_{k|\ell} = \mu_k + \theta_\ell + \epsilon_{k\ell} = \mu + \tau_k + \theta_\ell + \epsilon_{k\ell},
\end{equation}

\noindent in which $\mu_k$ denotes the mean performance of algorithm $a_k$ for the problem class $\Gamma$ after the (additive) effect of each instance, $\theta_\ell$, is blocked out; and $\epsilon_{k\ell}$ is the residual relative to that particular observation, i.e., the term that accounts for all other unmodelled effects that may be present (e.g., uncertainties in the estimation of $Y_{k|\ell}$). Notice that the algorithm mean $\mu_k$ can be further decomposed into a grand mean of all algorithms for the problem class of interest, $\mu$, and the effect of algorithm $a_k$ on this grand mean, represented by $\tau_k$. By construction, $\sum_{k=1}^{A}\tau_k = 0$.

As mentioned earlier, an usual approach for comparing multiple optimisation algorithms on multiple problem instances (which is an analogous problem to that of testing multiple classifiers using multiple data sets \cite{Demsar2006}) employs the usual 2-step approach provided in most introductory texts in statistics. It starts by using omnibus tests, such as CBD ANOVA or Friedman's test, to investigate the existence of some effect by testing hypotheses of the form:
\begin{equation}
\label{eq:hypotheses1}
\begin{split}
H_0: &~\tau_k = 0, \forall k\in\left[1,A\right];\\
H_1: &~\exists\tau_k \neq 0,
\end{split}
\end{equation}

\noindent which is equivalent to testing whether all algorithms have the same mean value \textit{versus} the existence of at least one algorithm with a different mean performance for the problem class of interest. If the omnibus test returns a statistically significant result at some particular confidence level, pairwise comparisons are then performed as a second step of inference, to pinpoint the significant difference(s). These are usually performed using common tests for the difference of means of two matched populations, such as variations of the paired t-test for post-ANOVA analysis, or Wilcoxon's Signed Ranks test (or the Binomial Sign Test) for post-Friedman analysis.

There are several different sets of pairwise comparisons that can be executed in this step, but the two most commonly used ones are \textit{all vs.~all} and the \textit{all vs.~one}. In the former, $A(A-1)/2$ tests are performed on the paired difference of means (again, with \textit{instance} as the pairing factor) for all pairs of algorithms $a_i\neq a_j$:
\begin{equation}
\label{eq:allVall}
\begin{split}
H_0^{(ij)}: &~\mu_{(ij)} = 0;\\
H_1^{(ij)}: &~\mu_{(ij)} \neq 0,
\end{split}
\end{equation}

\noindent in which $\mu_{(ij)} = \tau_i - \tau_j$ represents the mean of the paired differences in performance between $a_i$ and $a_j$. \textit{All vs.~all} pairwise comparisons are the default approach in the meta-heuristics literature, but in practice are only required when one is really interested in obtaining a complete ``ordering'' of all algorithms in a given experiment.

The second type of pairwise comparisons is the \textit{all vs.~one}, in which there is a reference algorithm (e.g. a new proposal one is interested in evaluating). Assuming that the first algorithm, $a_1$, is always set as the reference one, \textit{all vs.~one} pairwise comparisons are defined as:
\begin{equation}
\label{eq:allVone}
\begin{split}
H_0^{(j)}: &\mu_{(1j)} = 0;\\
H_1^{(j)}: &\mu_{(1j)} \neq 0,
\end{split}
\end{equation}

\noindent for all $j\neq 1$. This approach results in only $(A-1)$ hypotheses to be tested and, as we will discuss later, results in inferential procedures with higher statistical power or, if sample sizes are being estimated, in experiments requiring fewer instances to achieve a given desired power \cite{Mathews2010}.%
\footnote{Statistical power (or sample sizes) can be further improved if the alternative hypotheses in \eqref{eq:allVone} are directional,
\begin{equation*}
\begin{split}
H_0^{(j)}: &\mu_{(1j)} = 0;\\
H_1^{(j)}: &\mu_{(1j)} > 0,
\end{split}
\end{equation*}

\noindent which can be employed if the researcher is specifically interested in knowing whether or not the reference algorithm is superior (in terms of mean performance) to the others (i.e., if equality or inferiority are considered equally undesirable outcomes). Of course it makes no sense trying to use directional alternative hypotheses in the \textit{all vs.~all} case, since in that case there is no reference algorithm common to all tests.}

It is important to highlight at this point that the ANOVA or Friedman tests are not strictly required when performing the comparison of multiple algorithms. From an inferential perspective, it is equally valid to perform only the pairwise comparisons, as long as the rejection threshold of each hypothesis tested is adequately adjusted to prevent the inflation of Type-I error rates resulting from multiple hypothesis testing \cite{Shaffer1995}. While it is true that post-ANOVA pairwise tests can usually benefit from increased sensitivity (by using the more accurate estimate of the common residual variance that results from the ANOVA model), this improvement of statistical power is only achieved under the assumption of equality of variances, which is rarely true in algorithm comparisons. We argue that this potential (albeit occasional) disadvantage is more than compensated by a range of possible benefits that emerge when we focus on the pairwise comparisons:

\begin{itemize}
	\item Tests for the mean of paired differences are generally much simpler than their corresponding omnibus counterparts, and generally require fewer assumptions. For instance, unlike the CBD ANOVA, the paired t-test does not require equality of variances between algorithms, nor does it assume that the data from each algorithm is normally distributed -- only the paired differences need be approximately normal, and not even that if the sample size is large enough. These tests also yield much more easily interpretable results, and are considerably simpler to design and analyse.
	
	\item In the context of experimental comparison of algorithms one is commonly interested in relatively few tests: the number of algorithms compared is usually small, and in several cases tests can be performed using the \textit{all vs.~one} approach (e.g., when assessing the performance of a proposed method in comparison to several existing ones).  This alleviates the main issue with testing multiple hypotheses, namely the loss of statistical power (or, alternatively, the increase in the sample size required to achieve a certain sensitivity) that results from adjusting the significance levels to control the Type-I error rate. \cite{Shaffer1995}.
	
	\item From the perspective of estimating the number of required instances for algorithm comparisons, one of the most challenging quantities that need to be defined in the design stage of these experiments is a \textit{minimally relevant effect size} (MRES) \cite{Campelo2019}. The definition of MRES in the context of omnibus tests is very counter-intuitive, since it must be defined in terms, e.g., of the ratio of between-groups to within-groups variances. While still challenging, the definition and interpretation of MRES in the context of comparing a pair of algorithms is incomparably simpler: it essentially involves defining the smallest difference in mean performance between two algorithms (either in terms of raw difference or normalised by the standard deviation) that would have some practical consequence in the context of the research being performed \cite{Campelo2019}.
\end{itemize}

Given the considerations above we argue that, while it is common practice to perform the experimental analysis as the 2-step procedure outlined earlier, it is often more practical to focus on the pairwise tests, particularly from the perspective of sample size calculations. In the following sections we detail how the two sample sizes involved in the experimental comparison of algorithms -- number of instances, number of runs of each algorithm on each instance -- can be estimated by focusing on the pairwise comparisons. This of course does not preclude the use of omnibus tests as part of the analysis, or of any other type of analytical tool for investigating and characterising the behaviour and performance of the algorithms being tested. In fact, sample sizes estimated by focusing on the pairwise comparisons result in omnibus tests with at least the same statistical power for which the pairwise tests were planned, assuming that their assumptions, which are often more restrictive, are satisfied. However, the use of omnibus tests becomes optional, and the analyst can choose to proceed directly with the pairwise tests knowing that the statistical properties of their experiment are guaranteed. 

Finally, based on the statistical model from \eqref{eq:statmodel} we can define the two kinds of paired differences in performance, which we will examine in this work. The first one is the simple paired difference in performance, $D_{(ij)|\ell} = Y_{i|\ell} - Y_{j|\ell}$, which can be estimated from the data as 
\begin{equation}
\label{eq:estimate_simple}
\widehat{D}_{(ij)|\ell} = \bar{X}_{i\mid\ell} - \bar{X}_{j\mid\ell},
\end{equation}

\noindent in which $\bar{X}_{i\mid\ell}$ and $\bar{X}_{j\mid\ell}$ are the mean (or median) observed performance of algorithms $a_i$ and $a_j$ on instance $\gamma_\ell$, computed from $n_{i\ell}$ and $n_{j\ell}$ runs, respectively. An alternative approach is to use percent differences, which can be estimated for the \textit{all vs.~one} case as
\begin{equation}
\label{eq:estimate_perc1}
\widehat{D}_{(1j)|\ell} = \frac{\bar{X}_{1|\ell} - \bar{X}_{j|\ell}}{\bar{X}_{1|\ell}} = 1 - \frac{\bar{X}_{j|\ell}}{\bar{X}_{1|\ell}},
\end{equation}

\noindent and for the \textit{all vs.~all} case as
\begin{equation}
\label{eq:estimate_perc2}
\widehat{D}_{(ij)|\ell} = \frac{\bar{X}_{i|\ell} - \bar{X}_{j|\ell}}{\bar{X}_{\bullet|\ell}},
\end{equation}

\noindent in which $\bar{X}_{\bullet|\ell}$ is the estimate of the grand mean of all algorithms on instance $\ell$,
\begin{equation}
\label{eq:Xbardot}
\bar{X}_{\bullet|\ell} = \frac{1}{A}\sum\limits_{k=1}^{A}\bar{X}_{i\mid\ell}.
\end{equation}

\section{Calculating the number of repetitions for comparisons of multiple algorithms}
\label{sec:nruns}
\subsection{Preliminary Statistical Concepts}
Before we present the derivation of sample size formulas, there are a few statistical concepts that need to be clarified. More information and detailed discussions of these concepts are available in our previous work \cite{Campelo2019}, and here we will only present a very brief review of these definitions.

The {\it{minimally relevant effect size}} (MRES), which is an essential concept for determining the smallest sample size in the comparison of algorithm, is defined in the context of this work as \textit{the smallest difference between two algorithms that the researcher considers as having some practical effect}. There exists several different effect size estimators that are relevant for a variety of experimental questions \cite{Ellis2010}, but in the context of the hypotheses considered in this work we focus on Cohen's $d$ coefficient,
\begin{equation}
\label{eq:effect_size}
d = \frac{\delta}{\sigma} = \frac{\left|\mu_{ij}\right|}{\sigma},
\end{equation}

\noindent which is the (real) value of the mean of the paired differences\footnote{More generally, the denominator of Cohen's $d$ is the magnitude of the deviation between the actual value of a parameter and the value suggested under the null hypothesis. However, in this work we always assume the null hypothesis to suggest a difference of zero, hence the simpler definition.} normalised by $\sigma$, the (real) standard deviation of the paired differences. In the context of this work, the value in the numerator of \eqref{eq:effect_size} is estimated as the mean of the $|\widehat{D}|$ values calculated using \eqref{eq:estimate_simple} (for the simple paired difference) or \eqref{eq:estimate_perc1}--\eqref{eq:estimate_perc2} (for the percent paired differences).

Based on this definition, the MRES is defined as 
\begin{equation}
\label{eq:MRES}
d^* = \frac{|\delta^*|}{\sigma} = \frac{|\mu_{ij}^*|}{\sigma}
\end{equation}

\noindent i.e., as the smallest magnitude of the standardised mean of paired differences between two algorithms that would actually result in some effect of practical relevance.\footnote{Different areas have distinct standards of what constitutes small or large effect sizes in terms of $d$ \cite{Ellis2010}, and domain-specific knowledge needs to be considered when determining $d^*$. We provide some broad guidelines in our previous work \cite{Campelo2019}, but a broader investigation of generally accepted MRES values in the experimental research on meta-heuristics is yet to be conducted.}

Let $X_{k\mid\ell}^t$ denote the observation of performance of algorithm $a_k$ on the $t$-th run on a given instance $\gamma_\ell$, and 
\begin{equation}
\label{eq:xkbar}
\bar{X}_{k\mid\ell} = \frac{1}{n_{k\mid\ell}}\sum_{t = 1}^{n_{k\mid\ell}}X_{k\mid\ell}^t
\end{equation}

\noindent denote the sample estimator of the mean performance of $a_k$ on that instance, based on $n_{k\mid\ell}$ independent runs. Under relatively mild assumptions, we know that the sampling distribution of means converges to a Normal distribution even for reasonably small sample sizes, with
\begin{equation}
\label{eq:xkbar_distr}
\bar{X}_{k\mid\ell} \sim \mathcal{N}\left(\mu_{k\mid\ell}, \frac{\sigma^2_{k\mid\ell}}{n_{k\mid\ell}}\right).
\end{equation}

\noindent where the $\sigma^2_{k\mid\ell}~/~n_{k\mid\ell}$ term is the squared standard error of the sample mean. 

%Given a set of $A$ algorithms, let each performance observation $Y_{k\mid\ell}$ and each mean performance $\mu_k$ be decomposed as in (\ref{eq:statmodel})  and the definitions given in \ref{subsec:TH}.
%\begin{equation}
%\label{eq:effects_model}
%\mu_k = \mu + \tau_k,
%\end{equation}
%
%\noindent in which $\mu$ is the \textit{grand mean} of the performance of all $A$ algorithms on the instance, and $\tau_k$ represents the \textit{effect} of algorithm $a_k$, i.e.,  how much the mean performance of $a_k$ deviates from the grand mean. Note that, by construction, $\sum_{k=1}^A\tau_k = 0.$

%\subsection{Problem Statement}
Given a single problem instance $\gamma_\ell$ and the $A$ algorithms one wishes to compare, the approach for calculating the number of repetitions is similar to the one outlined in \cite{Campelo2019}, namely finding the smallest total number of runs such that the standard errors of estimation -- which can be interpreted as the \textit{measurement errors} of the pairwise differences in performance  -- can be controlled at a predefined level. This can be expressed as an optimisation problem,%
\begin{equation}
\label{eq:optprob}
\begin{split}
\mbox{Find }&\mathbf{n}_\ell^\star = \arg\min\sum_{k=1}^{A}n_{k\mid\ell}\\
\mbox{Subject to: }&se^2_{(ij)\mid\ell} \leq \left(se^*\right)^2,~\forall \left(a_i,a_j\right)~\mbox{pairs of interest}
\end{split}
\end{equation}

\noindent in which $n_{k\mid\ell}$ is the number of runs of algorithm $a_k$ on instance $\gamma_\ell$, $se_{(ij)\mid\ell}$ is the standard error of estimation of the relevant statistic (e.g., simple or percent difference of means) for a pair of algorithms $a_i,a_j$. The $\left(a_i,a_j\right)$ pairs of interest depend on the types of comparisons planned. While there can be a multitude of comparison types that can be performed, as said before, two comparisons are of particular interest in experimental algorithm: \textit{all vs.~all} and \textit{one vs.~all}. All pairs $\left(a_i,a_j\right):i\neq j$ are of interest in the \textit{all vs.~all} case while, In the \textit{all vs.~one} case, if $a_1$ is defined as the reference algorithm, then the pairs of interest are $\left(a_1,a_j\right): j\in\left[2,A\right]$. 

It is worth noting that the most adequate formulation for this problem is an integer one, but in this work we employ a relaxed version of the formulation that accepts continuous values. Since this formulation is used only to derive reference values for the optimal \textit{ratio} of sample sizes (which can be continuous even under the discrete sample size constraint) this relaxation does not result in any inconsistencies. Also notice that there is a second set of constraints that is omitted from formulation \eqref{eq:optprob}, namely $ n_{k\mid\ell} \geq 2,~\forall k\in\left[1,A\right]$, which are needed to guarantee that standard deviations can be estimated for each algorithm. We did not explicitly include these constraints since the proposed method assumes that all algorithms will be initially run a number $n_{0}>2$ times before iterative sample size estimations start, so these constraints will always be satisfied.

%\begin{itemize}
%	\item \textit{all vs.~all}, in which case one is interested in obtaining a complete picture of the relative performance of all algorithms involved. All pairs $\left(a_i,a_j\right):i\neq j$ are of interest in the \textit{all vs.~all} case.
%	\item \textit{all vs.~reference}, in which case there is a single reference (or control) algorithm that needs to be compared against all others. \textit{All vs.~one} appears whenever there is a single algorithm of greater interest (e.g., something that is being proposed) and one wants to investigate how it compares against well-established ones (e.g., state-of-the-art). In this case, if $a_1$ is defined as the reference algorithm, then the pairs of interest are $\left(a_1,a_j\right): j\in\left[2,A\right]$. 
%\end{itemize}

The specific calculation of the standard errors $se_{ij}$ depends on the type of differences being considered: simple differentes or percent differences and, in the latter case, whether the planned comparisons are an \textit{all vs.~one} or  \textit{all vs.~all} case. The results for the comparison of two algorithms were presented in our previous work \cite{Campelo2019}, and in this section we generalise those results to the case of comparisons of multiple algorithms. We also introduce a heuristic that can be used to estimate the numbers of repetitions in the case of more general statistics that may be of interest in the context of experimental comparison of algorithms.

\subsection{Derivation of optimal ratios $n_{i\mid\ell}/n_{j\mid\ell}$}
While it was possible to derive an analytic solution to the problem of determining the optimal ratio of sample sizes in the two algorithm case \cite{Campelo2019}, a similar closed solution does not seem to be possible for the general problem \eqref{eq:optprob} with $A \geq 3$. It is, however, possible to derive optimality-based ratios of sample sizes for any pair of algorithms, which allows us to propose a principled heuristic to generate adequate samples for a set of $A$ algorithms on any individual instance. The proposed heuristic iteratively generates observations of performance for each of the algorithms considered, so that the constraints $se^2_{(ij)\mid\ell} \leq \left(se^*\right)^2$ can be satisfied with as few total algorithm runs as possible. This heuristic is described in Algorithm \ref{alg:nruns}.

\begin{algorithm}
	\caption{Sample algorithms on a single instance.}
	\label{alg:nruns}
	\begin{algorithmic}[1]
		\Require Instance $\gamma_\ell$; Algorithms $a_1,\dotsc,a_A$; accuracy threshold $se^*$; initial sample size $n_0$; maximum sample size $n_{max}$.
		\State Generate $n_0$ initial samples of each algorithm on instance $\gamma_\ell$. 
		\State Estimate all $\widehat{se}_{ij}$ of interest
		\While{$\left(\exists~\widehat{se}_{(ij)\mid\ell} > se^*\right)$ and $\left(\sum_{k=1}^An_{k\mid\ell} < n_{max}\right)$}
		\State $\widehat{se}_{max}\leftarrow\underset{i,j}{\max}~\widehat{se}_{(ij)\mid\ell}$
		\State Perform \textbf{one} additional run of the algorithm that is contributing the most to $\widehat{se}_{max}$ on instance $\gamma_\ell$
		\State Estimate all $\widehat{se}_{(ij)\mid\ell}$ of interest.
		\EndWhile
		\vspace{.10cm}
		\State\Return $\mbf{x}_{k\mid\ell},~k=1,\dotsc,A$ \Comment{Vectors of observations generated for all algorithms.}
	\end{algorithmic}
\end{algorithm}

The procedure detailed in Algorithm \ref{alg:nruns} is general enough to accommodate any statistic one may wish to use for quantifying the pairwise differences of performance between algorithms, as long as the standard errors and their sensitivities to changes in $n_{k\mid\ell}$ can be estimated (which can be done using bootstrap, if analytical expressions are not available). In the case where analytical expressions for the optimal ratio of sample sizes between two algorithms are available (see below), the determination of the algorithm that is contributing the most to $\widehat{se}_{max}$ on a given instance $\gamma_\ell$ (line 5 of the algorithm) can be done based on these \textit{optimal ratios}: if the observed ratio is smaller than the optimal one, run the algorithm in the numerator; otherwise, run the one in the denominator.

Another point worth mentioning is that the procedure outlined in Algorithm \ref{alg:nruns} is essentially a greedy approach to reduce the worst-case standard error, which means that even if it is interrupted by the computational budget constraint the resulting standard errors of estimation will be the smallest ones achievable. This, in turn, means that the residual variance due to these estimation uncertainties will be as small as possible, which in turn increases the sensitivity of the experiment to smaller differences in performance by reducing, even if slightly, the denominator term in \eqref{eq:effect_size}.

The derivation of the standard errors and optimal ratios of sample sizes, for the simple and percent differences of means, both for the \textit{all vs.~one} and \textit{all vs.~all} cases, are presented below. Notice that while we can derive optimal sample size ratios for all paired differences of interest, Algorithm \ref{alg:nruns} will result in ratios that can deviate (sometimes substantially so) from the optimal reference values. This is a consequence of the fact that ratios of sample sizes are not independent quantities when multiple algorithms are compared. In practice, pairs that present the worst-case standard error ($\widehat{se}_{max}$) more often will tend to have ratios of sample sizes closer to the optimal values derived below, with the other pairs deviating from optimality due to the algorithm focusing the sampling on the pairs that are associated with $\widehat{se}_{max}$. This, however, does not reduce the validity and usefulness of the optimality derivations (as they are needed to decide which algorithm from the selected pair should receive a new observation), nor does it affect the quality of the results provided by Algorithm \ref{alg:nruns}.

Finally, for the sake of clarity we will omit the instance index $\ell$ from all derivations in the next section, but the reader is advised to keep in mind that the calculations in the remained of this section are related to a single given instance.

\subsubsection{Simple differences of means}

In the case of the simple differences of means, the statistic of interest for any pair of algorithms $a_i, a_j$ is
\begin{equation}
\label{eq:sdm}
\phi_{ij} = \mu_i - \mu_j = \tau_i - \tau_j,
\end{equation}

\noindent which is estimated using the sample quantities
\begin{equation}
\label{eq:sdme}
\widehat{\phi}_{ij} = \bar{X}_i - \bar{X}_j.
\end{equation}

Given \eqref{eq:xkbar_distr}, the sampling distribution of $\widehat{\phi}_{ij}$ will be
\begin{equation}
\label{eq:sdme_dist}
\widehat{\phi}_{ij} \sim \mathcal{N}\left(\mu_i-\mu_j,\sigma_i^2n_i^{-1} + \sigma_j^2n_j^{-1}\right)
\end{equation}

\noindent with squared standard error 
\begin{equation}
\label{eq:sdme_dist2}
se^2_{ij} = \sigma_i^2n_i^{-1} + \sigma_j^2n_j^{-1}.
\end{equation}

For any pair of algorithms $a_i,a_j$, the optimal sample size ratio $n_i/n_j$  for a given instance $\gamma$ can be found using the formulation in \eqref{eq:optprob}:
\begin{equation}
\label{eq:optprob_2algs}
\begin{split}
\mbox{Find }&\left[n_i,n_j\right]^\star = \arg\min f_{ij}\left(n_i,n_j\right) = n_i + n_j;\\
\mbox{Subject to: }&g_{ij}\left(n_i,n_j\right) = se^2_{ij} - \left(se^*\right)^2 \leq 0.
\end{split}
\end{equation}

\noindent This formulation has a known analytical solution \cite{Campelo2019} in terms of the ratio of sample sizes:
\begin{equation}
\label{eq:optratio-sdm}
r_{opt} = \frac{n_i^*}{n_j^*} = \frac{\sigma_i}{\sigma_j},
\end{equation}

\noindent which can be rewritten in terms of sample estimators as
\begin{equation}
\label{eq:optratioe-sdm}
\hat{r}_{opt} = \frac{S_i}{S_j},
\end{equation}

\noindent with $S_i, S_j$ being the sample standard deviations of the observations of algorithms $a_i, a_j$, respectively. In this case the decision rule for which algorithm should receive a new observation (line 5 of Algorithm \ref{alg:nruns}) can be expressed as: let $a_i, a_j$ be the two algorithms that define $\widehat{se}_{max}$. Then:
\begin{equation}
\label{eq:samplerule}
\begin{cases}
\mbox{If } n_i / n_j < \hat{r}_{opt} &\rightarrow \mbox{sample } a_i\\
\mbox{Else}  &\rightarrow \mbox{sample } a_j.
\end{cases}
\end{equation}

It is worthwhile to notice that this result is independent on whether one is interested in \textit{all vs.~all} or \textit{all vs.~one} comparisons, the only difference being which pairs $a_i,a_j$ generate $se_{ij}$ constraints in \eqref{eq:optprob}. 

\subsubsection{Percent differences of means}
\label{sec:percdiff}
For the percent differences of means\footnote{Strictly speaking the difference discussed here is the proportional (per unit) difference, which must be multiplied by a factor of $100$ to be actually expressed as \textit{percent}. This, however, has no effect in the derivations that follow, and is merely a matter of linear scaling.} we need to distinguish between the two types of comparisons discussed in this work. 

\paragraph{\textbf{All vs.~one}:} For \textit{all vs.~one} comparisons, assuming $a_1$ is the reference algorithm, the statistic of interest is given as:
\begin{equation}
\label{eq:pdm-1}
\phi_{1j} = \frac{\mu_1 - \mu_j}{\mu_1} = 1 - \frac{\mu_j}{\mu_1},
\end{equation}

\noindent which can be interpreted as the \textit{percent mean loss in performance of $a_j$ when compared to $a_1$}. This is estimated using
\begin{equation}
\label{eq:pdme-1}
\widehat{\phi}_{1j} = 1 - \frac{\bar{X}_j}{\bar{X}_1}, % = 1 + \frac{-\bar{X}_j}{\bar{X}_1},
\end{equation}

\noindent which is distributed according to one plus the ratio of two normal variables $R/S$, with
\begin{equation*}
\begin{split}
R &\sim \mathcal{N}\left(- \mu_j, \sigma_j^2n_j^{-1}\right);\\
S &\sim \mathcal{N}\left(\mu_1, \sigma_1^2n_1^{-1}\right).
\end{split}
\end{equation*}

If we can assume that all algorithms involved return observations that are strictly positive (which is quite common, e.g., in several families of problems in operations research as well as for several common performance indicators in multiobjective optimisation), the squared standard error of \eqref{eq:pdme-1} can be calculated as \cite{Campelo2019,Fieller1954,Franz2007}:
\begin{equation}
\label{eq:pdme_se1}
\begin{split}
se^2_{1j} &= \left(\frac{\mu_j}{\mu_1}\right)^2\left[\frac{\sigma_1^2n_1^{-1}}{\mu_1^2} + \frac{\sigma_j^2n_j^{-1}}{\mu_j^2}\right]\\
&= \frac{\sigma^2_1\mu_j^2}{\mu_1^4}n_1^{-1} + \frac{\sigma^2_j}{\mu_1^2}n_j^{-1}\\
&=C_1^{1j}n_1^{-1} + C_2^{1j}n_j^{-1},
\end{split}
\end{equation}

\noindent with 
\begin{equation*}
C_1^{1j} = \frac{\sigma^2_1\mu_j^2}{\mu_1^4}~~~~\mbox{and}~~~~C_2^{1j} = \frac{\sigma^2_j}{\mu_1^2}.
\end{equation*}

The KKT conditions for the problem formulation in \eqref{eq:optprob_2algs} state that, at the optimal point $\left[n_1^*, n_j^*\right]$ the following should be true:
\begin{equation}
\label{eq:KKT}
\begin{split}
\nabla f_{1j} + \beta_{1j}\nabla g_{1j} &= \mbf{0}\\
\beta_{1j}g_{1j} &= 0\\
\beta_{1j} &\geq 0
\end{split}
\end{equation}

The gradient of the objective function is trivially derived as $\nabla f_{1j} = \left[1,1\right]$, and the partial derivatives of $g_{1j}$ are:
\begin{equation}
\label{eq:dgdn-a}
\frac{\partial g_{1j}}{\partial n_1} = -C_1^{1j}n_i^{-2}~~~~\mbox{and}~~~~\frac{\partial g_{1j}}{\partial n_j} = -C_2^{1j}n_j^{-2},
\end{equation}

The first row of \eqref{eq:KKT} can then be expanded as:
\begin{equation}
\label{eq:pdm-opt1}
\begin{split}
\beta_{1j}C_1^{1j}n_1^{-2} &= 1\\
\beta_{1j}C_2^{1j}n_j^{-2} &= 1\\
\end{split}
\end{equation}

\noindent which means that $C_1^{1j}n_1^{-2} = C_2^{1j}n_j^{-2}$ at the optimal point.\footnote{Equations \eqref{eq:pdm-opt1} also mean that the Lagrange multiplier must be strictly greater than zero, which in turn means that $g_{1j}$ is active at the optimal point, as expected from a sample size minimisation perspective.} Isolating the ratio of sample sizes and simplifying finally yields:
\begin{equation}
\label{eq:pdm-opt2}
r_{opt} = \frac{n_1^*}{n_j^*} = \frac{\sigma_1/\mu_1}{\sigma_j/\mu_j},
\end{equation}

\noindent which can be estimated from the sample using:
\begin{equation}
\label{eq:pdm-ropt}
\hat{r}_{opt} = \frac{S_1}{S_j}\frac{\bar{X}_j}{\bar{X}_1}.
\end{equation}

Given this expression for $\hat{r}_{opt}$, the same rule expressed in \eqref{eq:samplerule} can be used in Line 5 of Algorithm \ref{alg:nruns}, simply replacing $n_i$ by $n_1$.

\vskip 2em

\paragraph{\textbf{All vs.~all:}} For \textit{all vs.~all} comparison using percent differences the question of which mean to use as the normalising term in the denominator must be decided. Assuming that no specific preference is given to any of the algorithms in this kind of comparison, we suggest using the grand mean $\mu$ instead of any specific mean $\mu_k$, i.e.:
\begin{equation}
\label{eq:pdm-3}
\phi_{ij} = \frac{\mu_i - \mu_j}{\mu}
\end{equation}

\noindent which can now be interpreted as the \textit{difference between the means of $a_i$ and $a_j$, in proportion to the grand mean}. This is estimated using
\begin{equation}
\label{eq:pdme-4}
\widehat{\phi}_{ij} = \frac{\bar{X}_i - \bar{X}_j}{\bar{X}_\bullet},
\end{equation}

\noindent where 
\begin{equation*}
\bar{X}_\bullet = \frac{1}{A}\sum_{k = 1}^A\bar{X}_k
\end{equation*}
is an estimator of the grand mean $\mu$. This estimator is unbiased regardless of unequal sample sizes,
\begin{equation}
E\left[\bar{X}_\bullet\right] = \frac{1}{A}\sum_{k = 1}^AE\left[\bar{X}_k\right] = \frac{1}{A}\sum_{k = 1}^A\left(\mu + \tau_k\right) = \mu + \frac{1}{A}\sum_{k = 1}^A\tau_k = \mu;
\end{equation}

\noindent and has a squared standard error of:
\begin{equation}
V\left[\bar{X}_\bullet\right] = \frac{1}{A^2}\sum_{k = 1}^AV\left[\bar{X}_k\right] = \frac{1}{A^2}\sum_{k = 1}^A\sigma^2_kn_k^{-1},
\end{equation}

Under the same mild assumptions under which $\bar{X}_k \sim \mathcal{N}\left(\mu_k,\sigma_k^2n_k^{-1}\right)$, we have that the sampling distribution of this estimator is progressively closer to a normal distribution as the sample sizes are increased. These considerations allow us to calculate the standard error of \eqref{eq:pdme-4} in the same manner as \eqref{eq:pdme_se1}, using the simplified formulation of Fieller's standard error \cite{Fieller1954,Franz2007}, i.e.:
\begin{equation}
\label{eq:pdme_se2}
\begin{split}
se^2_{ij} &= \left(\frac{\mu_i - \mu_j}{\mu}\right)^2\left[\frac{\sigma_i^2n_i^{-1} + \sigma_j^2n_j^{-1}}{\left(\mu_i - \mu_j\right)^2} + \frac{\sum_{k = 1}^A\sigma^2_kn_k^{-1}}{A^2\mu^2}\right]\\
&=C_1^{ij}\left(\sigma_i^2n_i^{-1} + \sigma_j^2n_j^{-1}\right) + C_2^{ij},
\end{split}
\end{equation}

\noindent with 
\begin{equation*}
\begin{split}
C_1^{ij} &= \frac{1}{\mu^2} + \left(\frac{\mu_i - \mu_j}{A\mu^2}\right)^2 = \frac{1}{\mu^2}\left(1 + \frac{\phi^2_{ij}}{A^2}\right);\\
C_2^{ij} &= \left(\frac{\mu_i - \mu_j}{A\mu^2}\right)^2\sum_{\substack{k=1\\k\neq i,j}}^A\sigma^2_kn_k^{-1} =
\frac{1}{\mu^2}\frac{\phi^2_{ij}}{A^2}\sum_{\substack{k=1\\k\neq i,j}}^A\sigma^2_kn_k^{-1},
\end{split}
\end{equation*}

\noindent with $\phi_{ij}$ defined as in \eqref{eq:pdm-3}. This new expression for the standard error leads to \eqref{eq:dgdn-a} being replaced by:
\begin{equation}
\label{eq:dgdn-b}
\frac{\partial g_{ij}}{\partial n_1} = -C_1^{ij}{\sigma_i}^2 n_i^{-2},~~~~~~~~~~~~\frac{\partial g_{ij}}{\partial n_j} = -C_1^{ij}{\sigma_j}^2n_j^{-2},
\end{equation}

\noindent which, when inserted into the KKT conditions \eqref{eq:KKT} yields:
\begin{equation}
\label{eq:pdm-opt3}
\sigma_i^2n_1^{-2} = \sigma_j^2n_j^{-2}
\end{equation}

\noindent and, consequently, 
\begin{equation}
\label{eq:pdm-ropt2}
r_{opt} = \frac{n_i^*}{n_j^*} = \frac{\sigma_i}{\sigma_j},
\end{equation}

\noindent which is identical to the optimal ratio for the simple differences of means, and can be similarly estimated using the sample standard deviations. Once again, the same decision rule from \eqref{eq:samplerule} can be used.

\section{Calculating the number of required instances for comparisons of multiple algorithms }
\label{sec:ninstances}
In this section we deal with the calculation of the number of instances, $N$, required to obtain an experiment with predefined statistical properties. In a previous work \cite{Campelo2019} we developed sample size formulas for the comparison of two algorithms on a problem class of interest. Since we have reduced the multiple algorithm comparison problem to a series of pairwise comparisons (at least for the purposes of sampling), most of the theoretical justifications remain the same, and the only actual difference in the calculation of the required number of instances is the need to adjust the significance levels so as to maintain the overall probability of Type-I errors (false positives) controlled at the desired level $\alpha$ \cite{Shaffer1995}. 

The calculation of the number of instances for comparing two algorithms is based on the definition of a few desired statistical properties for the test \cite{Campelo2019}. More specifically, we want the test to have a predefined statistical power, $\pi^* = (1-\beta^*)$, to detect differences equal to or greater than a \textit{minimally relevant effect size} $d^*$, at a predefined significance level $\alpha$. This led to the definition of the smallest required number of instances for the comparison of two algorithms (based on the paired t-test, for a two-sided alternative hypothesis) as:
\begin{equation}
\label{eq:calcN}
N^* = \min N \mid t_{1-\alpha/2}^{(N-1)}\leq t_{\beta^*; \left|ncp^*\right|}^{(N-1)}
\end{equation}

\noindent in which $t_{q}^{(N-1)}$ is the $q$-quantile of the central t distribution with $N-1$ degrees of freedom, and $t_{q; \left|ncp^*\right|}^{(N-1)}$ is the $q$-quantile of the non-central t distribution with $N-1$ degrees of freedom and noncentrality parameter $ \left|ncp^*\right| = \left|d^*\right|\sqrt{N}$. The required sample sizes can be further reduced in the \textit{all vs.~one} comparisons if one-sided alternative hypotheses can be used, in which case we simply replace $t_{1-\alpha/2}^{(N-1)}$ by $t_{1-\alpha}^{(N-1)}$ in \eqref{eq:calcN}.\footnote{The derivation of these results and a review of the statistical concepts associated with it are available in \cite{Campelo2019}. In that work we also showed how these results can be easily extended to the design of experiments based on nonparametric alternatives to the paired t-test, such as Wilcoxon's Signed Ranks test or the Binomial Sign test, by using the asymptotic relative efficiency of these tests \cite{Montgomery2013,Sheskin2011}. The same generalisations apply to the current work.}  

Since our proposed protocol for sample size calculation is based on designing the experiments from the perspective of the pairwise tests that will need to be performed, the sample size formulas outlined above remain valid, and we only need to adjust the significance level to account for the testing of multiple hypotheses.  As mentioned previously, we have two common cases in the experimental comparison of algorithms: \textit{all vs.~all} comparisons  \eqref{eq:allVall}, which result in $K = A(A-1)/2$ tests; and \textit{all vs.~one} comparisons \eqref{eq:allVone}, in which case only $K = A-1$ comparisons are performed. 

As discussed by Juliet Shaffer in her review of multiple hypothesis testing  \cite{Shaffer1995}, ``\textit{when many hypotheses are tested, and each test has a specified Type I error probability, the probability that at least some Type I errors are committed increases, often sharply, with the number of hypotheses}''. This is quite easy to illustrate if we consider that each test has a base probability of falsely rejecting a true null hypothesis given by its significance level $\alpha$. If $K$ comparisons are performed, and assuming that (i) the comparisons are independent, and (ii) all null hypotheses are true, the overall probability of observing one or more false positives (commonly called the \textit{family-wise error rate}, or FWER) can be calculated as
\begin{equation*}
\begin{split}
FWER = P\left(fp\geq 1\right) &= 1 - P\left(fp = 0\right)\\
&= 1 - \left(1-\alpha\right)^K.
\end{split}
\end{equation*}

As an example, if we consider a somewhat typical case of \textit{all vs.~all} comparisons of 5 algorithms (which would generate $K = 10$ hypotheses) with a per-comparison $\alpha = 0.05$, this would result in an $FWER = 1 - 0.95^{10} = 0.401$, i.e., a chance of around $40\%$ of falsely rejecting at least one of the null hypotheses. 

\begin{figure}[ht]
	\centering
	\includegraphics[width = \linewidth]{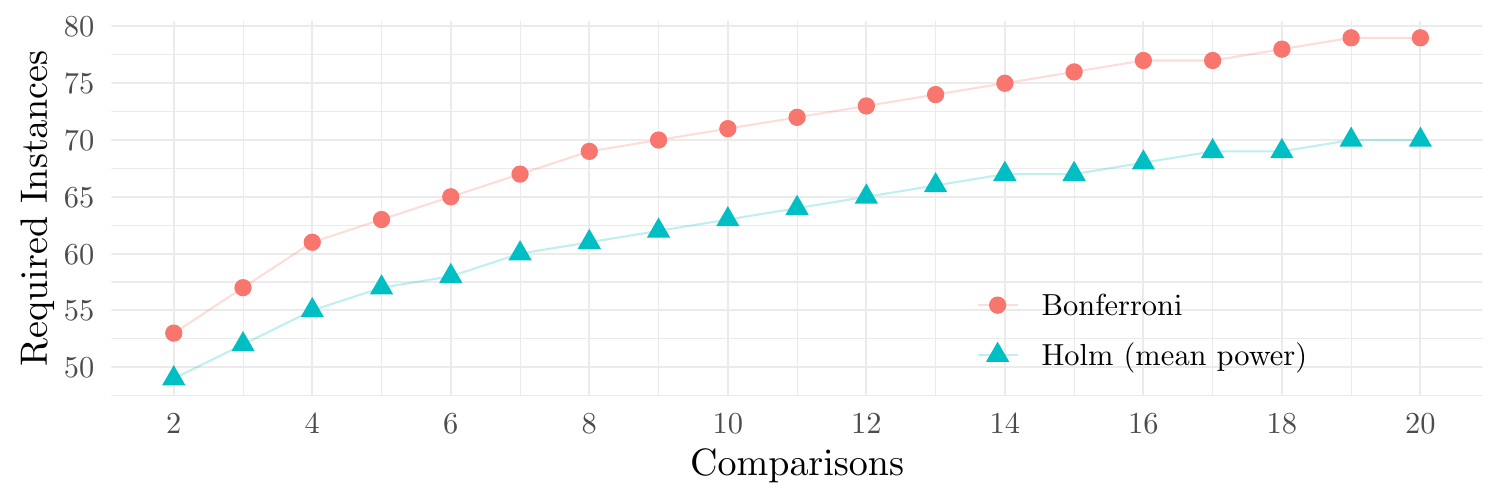}
%	\resizebox{1.0\linewidth}{!}{\input{HolmVsBonf.tex}}
	\caption{Example of required sample sizes for Bonferroni-corrected tests in a common case with $\alpha_f = 0.05,\pi^*=0.9$ and $d^* = 0.5$, considering a two-sided alternative, in comparison with the sample sizes required for obtaining the same mean power under Holm's correction. For reference, fifteen comparisons would represent \textit{all vs.~all} comparisons involving six algorithms. }
	\label{fig:BonferroniSS}
\end{figure}

To prevent this inflation of the FWER, a wide variety of techniques for multiple hypothesis testing are available in the literature. Possibly the most widely known is the Bonferroni correction \cite{Dunn1961,Shaffer1995}, which can maintain the FWER strictly under a desired level $\alpha_{f}$ by using, for each comparison, a reduced significance level $\alpha^\prime = \alpha_f / K$ (or, equivalently, by multiplying all p-values by $K$ and comparing against the original $\alpha$). This correction is known to be overly conservative, leading to substantial reductions of the statistical power of each comparison,  or to substantially higher sample sizes being required to achieve the same power. However, its simplicity often makes it quite appealing, particularly when few hypotheses need to be tested, in which case the penalisation applied to $\alpha$ is not too extreme. If the Bonferroni correction is selected to control the FWER in a given experiment, the calculation of the number of instances $N$ can be performed by simply dividing the value of $\alpha$ in \eqref{eq:calcN} by the number of planned comparisons ($K = A-1$ for the \textit{all vs.~one} case, or $K = A(A-1)/2$ for \textit{all vs.~all}). Figure \ref{fig:BonferroniSS} illustrates an example of the increase in the required sample size for Bonferroni-corrected tests as the number of comparisons is increased.

A uniformly more powerful alternative that also controls the FWER at a predefined level is Holm's step-down method \cite{Holm1979,Shaffer1995}. Instead of testing all hypotheses using the same corrected significance level, Holm's method employs a sequential correction approach. First, the p-values for all tests are computed. The hypotheses and their corresponding p-values are then ranked in increasing order of p-values, such that $p_1 \leq p_2 \dotsc \leq p_K$. Each of these ordered hypotheses is then tested at a different, increasingly stricter significance level,
\begin{equation}
\label{eq:holm}
\alpha^\prime_r = \alpha_f / \left(K - r + 1\right)
\end{equation}

\noindent in which $r$ is the rank of the hypothesis being tested. If $R$ is the largest value of $r$ for which a hypothesis would be rejected at the corresponding $\alpha^\prime_r$ level, then Holm's step-down method leads to the rejection of all null hypotheses with ranks $r\leq R$. Holm's method is also known to maintain the FWER under the desired value \cite{Shaffer1995}, but it leads to significantly less conservative tests than the Bonferroni correction. However, while the calculation of the number of instances is quite straightforward for the Bonferroni correction, it requires some clarification when Holm's method is employed. This is because each test is performed using a different significance level, which leads to individual tests with heterogeneous power levels under a constant sample size. 

One possible approach, which is found in the statistical literature, is to calculate the sample size based on the least favourable condition. We can calculate the sample size such that the test with the most strict significance level $\alpha^\prime_r$ will have at least the desired power $\pi^*$, which guarantees that the statistical power for all other comparisons will be greater than $\pi^*$. A simple examination of \eqref{eq:holm} shows that the most strict test will occur for $r = 1$, in which case $\alpha^\prime_r = \alpha_f/K$, i.e., the same corrected significance level generated by the Bonferroni correction. In this case, the same simple adjustment can be made for the formulas in \eqref{eq:calcN}, namely dividing the value of $\alpha$ by $K$. A related possibility, which generalises this one, is to calculate the sample size so that some predefined number of tests can have a statistical power of at least $\pi^*$. This can also be easily set up by determining a number $K^\prime$ of comparisons that should have power levels above the predefined threshold, and then divide $\alpha$ by $K - K^\prime + 1$ in \eqref{eq:calcN}. 

A second possibility, which is also relatively straightforward, is to design the experiment so that the mean (or median) power of the tests is maintained at the nominal level. The median case is roughly equivalent to setting $K^\prime = \lceil K/2\rceil$ and applying the method outlined above. Estimating sample size for achieving a mean power of $\pi^*$ is more challenging from an analytic perspective, but can be done iteratively without much effort using Algorithm \ref{algo:meanpower}. In this pseudocode $\mbox{\texttt{SampleSize}}\left(\alpha,\pi,d^*,H_1\right)$ denotes the calculation of the required sample size for performing a hypothesis test (e.g., using a paired t-test) with significance level $\alpha$, power $\beta$, MRES $d^*$ and type of alternative hypothesis $H_1$, as defined in \eqref{eq:calcN}. Similarly,  $\mbox{\texttt{Power}}\left(\alpha,n,d^*,H_1\right)$ calculates the power of a test procedure under the same parameters, assuming a sample size of $n$. Notice that since power increases monotonically with $\alpha$, the final sample size returned by the procedure outlined in Algorithm \ref{algo:meanpower} is guaranteed to be smaller than the result based on the least favourable condition.
\begin{algorithm}
	\caption{Estimate sample size for mean power in Holm's procedure.}
	\label{algo:meanpower}
	\begin{algorithmic}[1]
		\Require Desired mean power $\left(\pi^*\right)$; desired FWER $\left(\alpha_f\right)$; type of alternative hypotheses (one or two-sided) $\left(H_1\right)$; number of comparisons ($K$)
		\State $\bar{p} \leftarrow 0$
		\State $N\leftarrow\mbox{\texttt{SampleSize}}\left(\alpha_f,\pi,d^*,H_1\right) - 1$ \Comment{Using \eqref{eq:calcN}}
		\While{$\bar{p} < \pi^*$}
		\State $N \leftarrow N+1$
		\For{$i \in \left\{1,\dotsc,K\right\}$}
		\State $p_i \leftarrow\mbox{\texttt{Power}}\left(\alpha/i,N,d^*,H_1\right)$ \Comment{See \cite{Campelo2019} for details.}
		\EndFor
		\State $\bar{p} = \sum_{i=1}^{K}p_i/K$		\Comment{Calculate mean power}
		\EndWhile
		\vspace{.15cm}
		\State\Return $N$\;
	\end{algorithmic}
\end{algorithm}

\begin{figure}[ht]
	\centering
	\includegraphics[width = \linewidth]{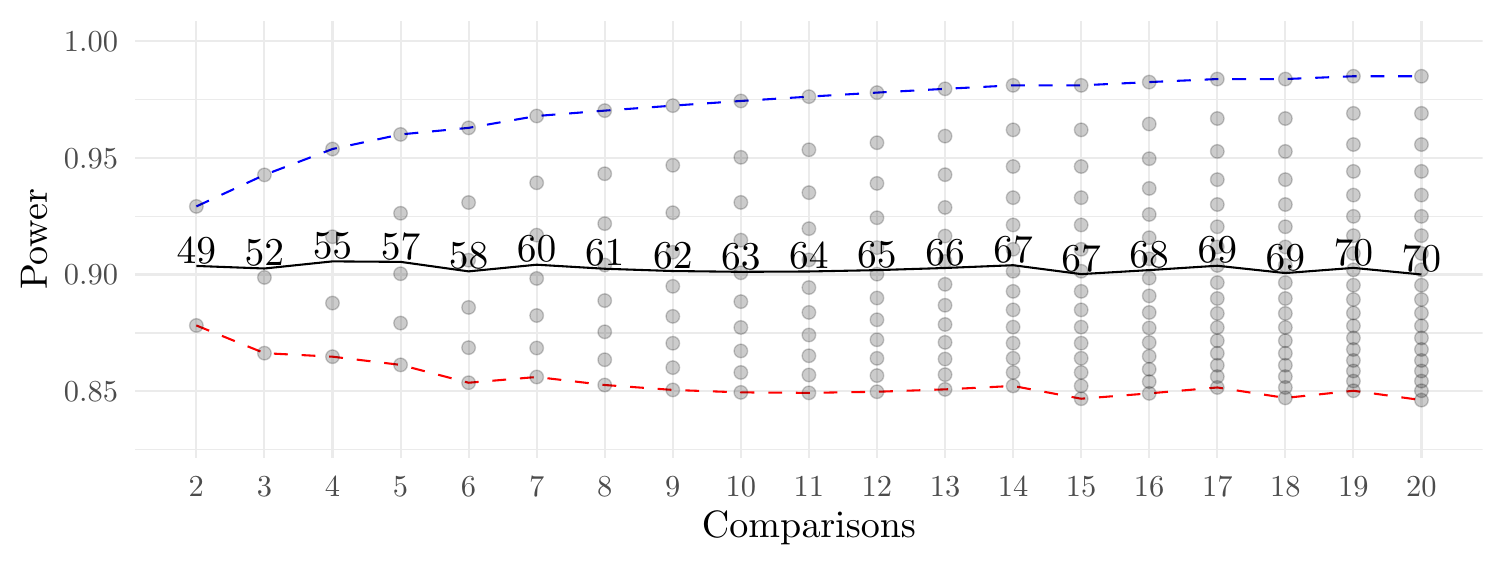}
	%\resizebox{1.0\linewidth}{!}{\input{HolmPower.tex}}
	\caption{Example of the resulting power of each comparison when designing experiments focused on mean power using Holm's method, in a common case with $\alpha_f = 0.05,\pi^*=0.9$ and $d^* = 0.5$, considering a two-sided alternative. The lines illustrate the mean, best and worst-case power values, and the gray dots represent the power of each comparison. The labels following the mean-power line inform the sample sizes required for each value of $K$. Notice that the worst-case power seems to stabilize around $0.85$ in this particular case.}
	\label{fig:HolmPower}
\end{figure}

Figure \ref{fig:HolmPower} illustrates the resulting power profile of an experiment designed for mean power using the procedure outlined in Algorithm \ref{algo:meanpower}. Two interesting aspects of this planning are immediately clear. First, the worst-case power appears to stabilize at about $0.85$, which is not much lower than the desired $0.9$, and would still be considered an acceptable power level in most applications. The second one is that the resulting sample sizes are, as expected, lower than those required if the Bonferroni correction were to be used. 

Based on these considerations, we can offer some suggestions as to how to proceed calculating the required number of instances ($N$) for the comparison of multiple algorithms using multiple problem instances:
\begin{itemize}
	\item It is more efficient, often considerably so, to use Holm's correction instead of Bonferroni's, given that the former requires smaller sample sizes to achieve similar power characteristics (see Figure \ref{fig:BonferroniSS}). 
	\item In general it seems reasonable to design experiments based on average power for Holm-corrected tests, given that the resulting worst-case power will not be much lower than the nominal value $\pi^*$, and the number of instances is substantially lower than what would be required for Bonferroni-corrected tests (see Figure \ref{fig:HolmPower}).
	\item If worst-case power must be guaranteed, the number of instances can be estimated using the formulas for the Bonferroni case, but the actual analysis should still be conducted using Holm's method, which in this case will result in better statistical power for all comparisons.
	\item Whenever possible, \textit{all vs.~one} comparisons should be used instead of \textit{all vs.~all}, since they result in fewer hypotheses being tested and, consequently, smaller sample sizes (or larger power if sample sizes are constant).
	\item If one-sided alternative hypotheses make sense in the experimental context of \textit{all vs.~one} comparisons they should also be preferred, since this also improves efficiency (in terms of requiring fewer instances to achieve a predefined power).
\end{itemize}

It is worth repeating here an important point also highlighted in our previous work \cite{Campelo2019}: the sample sizes calculated with the methodology presented in this section represent the \textit{smallest} sample size required for a given experiment to present desired statistical properties. This can be particularly useful when, e.g., designing new benchmark sets, sampling a portion of an existing (possibly much larger) set of test functions as part of algorithm development, or designing experiments in computationally expensive contexts. If more instances are available in a particular experiment they can of course be used, which will result in experiments with even higher statistical power and can enable the execution of finer analyses, e.g., to investigate the effect of dimension or other instance characteristics on performance. However, even if that is the case, we argue that the proposed methodology can still be useful for a number of reasons:

\begin{itemize}
	\item It can be used by researchers to obtain a more solid understanding of the statistical properties of their experiments (e.g., by examining power $\times$ effect size curves at a given sample size).
	\item By defining a MRES \textit{a priori}, the proposed approach provides a sanity check for researchers in overpowered experiments, with can result in arbitrarily-low p-values even for minuscule differences of no practical consequence \cite{Mathews2010}. That is, it provides a \textit{practical relevance} aspect, to contrast with the \textit{statistical significance} of the results.
	\item Even if the calculation of the number of instances is considered unnecessary in the context of a given experiment, the methodology for estimating the number of repeated runs of each algorithm on each instance can still be used to provide a statistically principled way to generate the data.
\end{itemize}

In the next section we provide an example of application of the proposed methodology for investigating the contribution of different neighbourhood structures for a scheduling problem. Please notice, however, that the main objective of the experiment is to demonstrate the capabilities of the proposed methodology for defining the sample sizes, and not necessarily to answer specific questions regarding the algorithms used in the example.

\section{Application Example}
\label{sec:application}
\subsection{Problem Description}
In the \textit{unrelated parallel machine scheduling problem (UPMSP) with sequence dependent setup times} \cite{Graham1979, Lawler1993,Vallada2011} a number of jobs, $J$, needs to be processed by a number of machines, $M$, minimising the completion time of the last job to leave the system, i.e., the makespan. This scheduling problem also presents a few other specificities which are not relevant for the current discussion, but are presented in detail in the relevant literature \cite{Vallada2011,Santos2016}.

So far the best algorithm for the solution of this problem seems to be a finely-tuned Simulated Annealing (SA) proposed by Santos \textit{et al.} in 2016 \cite{Santos2016}. Santos' SA was tested on a benchmark set originally introduced by Vallada and Ruiz \cite{Vallada2011}, which is composed of 200 tuning instances with $J\in\{50,100,150,200,250\}$ jobs and $M\in\{10,15,20,25,30\}$ machines. For each $\left(M, J\right)$ pair, eight instances with different characteristics (e.g., distribution of setup times) are present. A larger set, composed of 1000 instances, is also available to test hypotheses derived using the tuning set.

One of the core aspects of Santos' SA is the use of six neighbourhood structures to explore the performance landscape of the problem, namely \textit{Shift}, \textit{Two-shift}, \textit{Task Move}, \textit{Swap}, \textit{Direct Swap}, and \textit{Switch}. At each iteration one of these perturbation functions is selected randomly and generates a new candidate The authors do not provide any discussion related to how much each of these perturbation functions affects the performance of the algorithm. However, preliminary results \cite{Maravilha2019, Pereira2019} suggest that the six neighbourhood structures have considerably different effects on the algorithm, with \textit{Task move} having a much more critical influence than the others. 

\subsection{Experimental questions}
To further investigate the effect of the different neighbourhoods on the performance of the algorithm, we designed an experiment in which the \textit{full} algorithm (i.e., equipped with all six neighbourhood structures) was compared against variants with perturbation functions suppressed from the pool of available movements. We focussed on removing at most two perturbations from the pool, resulting in 21 algorithm variants (6 without a single perturbation, and 15 without two perturbations) which enabled the quantification of contributions of neighbourhood structures to the algorithm's ability to navigate the performance landscape of this particular problem class.

Given that we wanted to investigate the effects of removing these neighbourhood functions from the \textit{full} algorithm, an \textit{all vs.~one} design was be the most appropriate, resulting in a total of 21 hypotheses to be tested. Also, the variability in instance hardness suggested the use of percent differences (see Section \ref{sec:percdiff}), which quantify differences in a scale that is usually more consistent across heterogeneous instances.

\subsection{Experimental parameters}
We set the MRES at a (reasonably high) value of $d^* = 0.5$, i.e., we were interested in detecting variations in the mean of percent differences of at least half a standard deviation. We decided to design our experiment with a mean power $\pi^* = 0.8$, under a familywise significance level $\alpha = 0.05$ using Holm's correction. To prevent excess variability in the estimates of within-instance differences from inflating the overall residual variance of the experiment, we set the desired accuracy for the estimates as $se^* = 0.05$, i.e., a $5\%$ error in the estimates of paired percent differences on each instance. This relatively good accuracy can also be useful if we later want to further investigate the differences on individual instances, or even instance subgroups.

The method described in Section \ref{sec:ninstances} yielded $N = 57$ instances as the smallest amount necessary to obtain an experiment with the desired properties.\footnote{The full methodology for calculating the number of instances and algorithm runs is implemented in the \texttt{R} package \texttt{CAISEr} version 1.0.5, available at \url{https://cran.r-project.org/package=CAISEr}. The Java implementation of Santos' algorithm, as well as the instances used, can be retrieved online from \url{https://github.com/fcampelo/upmsp}.} At this point we could have opted to use any number of instances greater than or equal to this value, up to the full available set of 200 \cite{Vallada2011}. We opted to use only the suggested value of 57 not only because it was enough to guarantee the desired statistical properties for our experiment, but also to maintain a reserve of ``untouched'' tuning instances, which could be useful for further testing of eventual proposals of algorithmic improvement.\footnote{This is also a way to safeguard against overfitting our algorithms to the benchmark set, which is a common problem in algorithm design for problems with standard benchmark sets, such as the one being considered here \cite{BartzBeielstein2015,Hooker1994,Hooker1996}.}

The $A = 22$ algorithms (\textit{full} algorithm plus the 21 variants) were sampled on each instance under a maximum budget of $50A = 50\times 22 = 1100$ runs per instance, using an initial sampling of $n_0 = 10$ runs/algorithm/instance.

\subsection{Experimental results}
\begin{landscape}
\begin{figure}[ht]
	\centering
	\includegraphics[width = \linewidth]{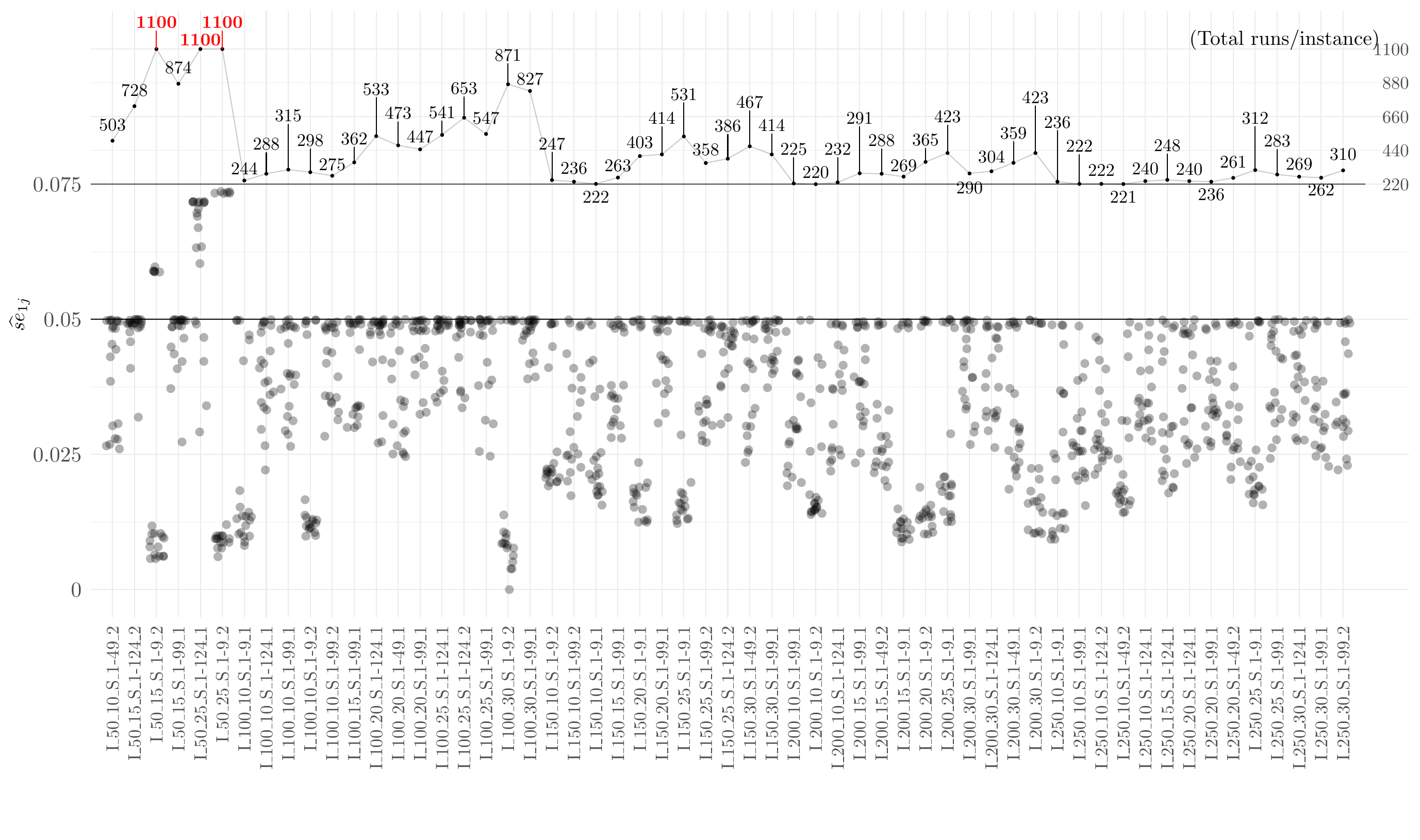}
	%\resizebox{\linewidth}{!}{\input{Exp1_SEandSS.tex}}
	\caption{Standard errors of estimation of the percent differences in performance between the \textit{full} algorithm and each of the 21 variants, for all instances tested. The numbers at the top indicate the total number of runs performed on each instance, and the labels on the x-axis provide the instance identifiers. Notice that only in three cases the allocated budget was insufficient to reduce all standard errors below the preset limit of $se^* = 0.05$.}
	\label{fig:Exp1SESS}
\end{figure}
\end{landscape}

Figure \ref{fig:Exp1SESS} shows the standard errors obtained for all algorithm pairs, as well as the total sample sizes generated, for each instance used in this experiment. It is very clear that the proposed approach for calculating the number of runs was able to control the standard errors at the desired level of $se^* = 0.05$ using, in most cases, a fraction of the total available computational budget. In only three instances was the available budget insufficient to bring all standard errors under the desired threshold, but even in these cases the resulting standard errors were controlled at relatively low values. These three particular instances all had $J = 50$ jobs (one with $M = 15$ and two with $M = 25$ machines), which suggests further investigation into the behaviour of the methods tested on this particular subset of instance sizes. 

\begin{figure}[h]
	\centering
	\includegraphics[width = \linewidth]{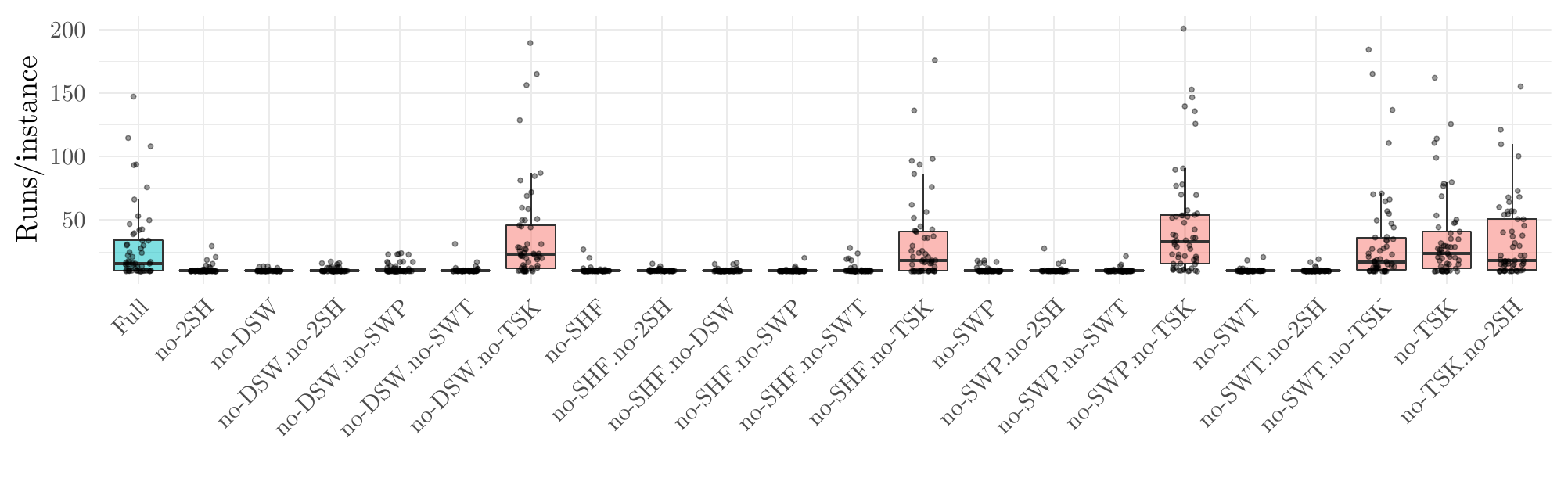}
	%\resizebox{\linewidth}{!}{\input{Exp1_nij.tex}}
	\caption{Distribution of $n_{k\mid\ell}$, i.e., of the number of runs for each algorithm variant on each instance. The majority of runs were allocated either to the \textit{full} algorithm - which was involved in all comparisons - or to variants that had the \textit{Task Move} (TSK) suppressed, suggesting a strong effect of this perturbation function.}
	\label{fig:Exp1Nij}
\end{figure}

Figure \ref{fig:Exp1Nij} shows the distribution of sample sizes for each algorithm involved in this experiment. As expected, the \textit{full} algorithm received a reasonably large number of observations for the majority of instances, since it was involved in all comparisons. Another feature that becomes quite clear in this figure is that all versions in which the \textit{Task Move} (TSK) neighbourhood function was suppressed also received much more observations, which suggests this neighbourhood as strongly influential and corroborates preliminary results obtained for Santos' SA on the UPMSP \cite{Pereira2019,Maravilha2019}.

\begin{figure}[ht]
	\centering
	\includegraphics[width = \linewidth]{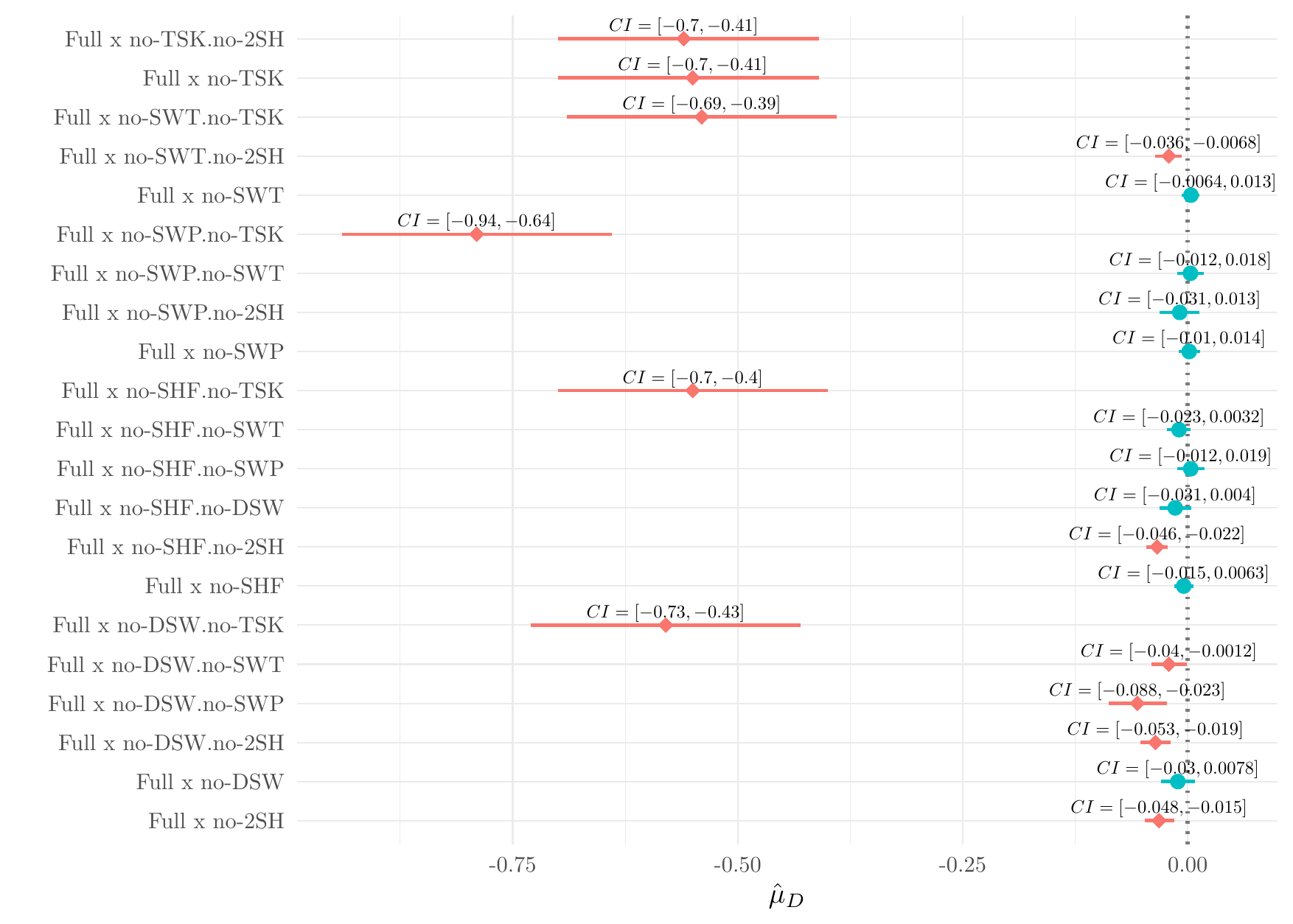}
	%\resizebox{\linewidth}{!}{\input{Exp1_CIs.tex}}
	\caption{Point estimates and confidence intervals (95\% joint confidence level) for the mean of paired percent differences of performance for the 21 comparisons of algorithm variants against the \textit{full} simulated annealing. Red intervals with a diamond marker indicate statistically significant results, while green intervals with circle markers indicate those where the null hypotheses were not rejected. Apart from the variants without the \textit{Task Move} perturbation function (TSK), all others resulted only in minor performance degradation, if any.}
	\label{fig:Exp1CI}
\end{figure}

The results of the pairwise comparisons are shown in Figure \ref{fig:Exp1CI}, which provides the confidence intervals (adjusted for a familywise error rate of $\alpha = 0.05$) for all comparisons between the \textit{full} algorithm and the suppressed variants. As indicated by the relative importance attributed to the methods by the sampling method, the suppression of the TSK neighbourhood function led to strong degradations of performance, with mean percent losses around $50\%$ or more. Also of interest in this table is the last column, which shows that the desired statistical properties of the experiment - particularly the ability to detect effects at or above approximately $d^* = 0.5$ - were actually obtained in the designed experiment.

Of possibly equal interest in terms of understanding the contributing factors to algorithm performance is the fact that the algorithm did not experience significant performance losses after the removal of some of its neighbourhood structures. Table \ref{tab:exp1res} presents the results of the t-tests\footnote{Graphical analysis of residuals indicated that the sampling distributions of means in all cases were sufficiently close to normality, allowing the use of t-tests for inference.} on the mean of paired percent differences of performance. Examining the bottom rows of this table, it shows that the suppression of \textit{Swap} (SWP), \textit{Switch} (SWT) and \textit{Shift} (SHF), individually as well as in pairs, resulted in no significant differences in performance. The width of the associated confidence intervals also suggests that any effects that may have gone undetected must be rather small.

\renewcommand{\arraystretch}{1.2}
\begin{table}[ht]
	\centering\footnotesize
	\caption{Summary of the 21 comparisons against the \textit{Full} algorithm, using Holm's step-down procedure.}
	\begin{tabular}{llllr}
		\hline
		Comparison & $\alpha_r^\prime$ & p-value & Confidence interval & $\widehat{d}~~$ \\
		\hline
		Full $\times$ no-SWP.no-TSK & $0.0024$ & $3.7\times 10^{-23}$ & $-0.79\pm 0.15$  & $-2.2$ \\
		Full $\times$ no-TSK.no-2SH & $0.0025$ & $1.7\times 10^{-17}$ & $-0.56\pm 0.14$ & $-1.6$ \\
		Full $\times$ no-DSW.no-TSK & $0.0026$ & $5.2\times 10^{-17}$ & $-0.58\pm 0.15$ & $-1.6$ \\ 
		Full $\times$ no-TSK & $0.0028$ & $1.0\times 10^{-16}$ & $-0.55\pm 0.15$ & $-1.6$ \\
		Full $\times$ no-SWT.no-TSK & $0.0029$ & $4.4\times 10^{-16}$ & $-0.54\pm 0.15$ & $-1.5$ \\
		Full $\times$ no-SHF.no-TSK & $0.0031$ & $6.3\times 10^{-16}$ & $-0.55\pm 0.15$ & $-1.5$ \\
		Full $\times$ no-SHF.no-2SH & $0.0033$ & $3.6\times 10^{-12}$ & $-0.034\pm 0.012$ & $-1.2$ \\
		Full $\times$ no-DSW.no-2SH & $0.0036$ & $3.5\times 10^{-8}$ & $-0.036\pm 0.017$ & $-0.85$ \\
		Full $\times$ no-2SH & $0.0038$ & $3.3\times 10^{-7}$ & $-0.032\pm 0.016$ & $-0.77$ \\
		Full $\times$ no-DSW.no-SWP & $0.0042$ & $4.5\times 10^{-6}$ & $-0.056\pm 0.033$ & $-0.67$ \\
		Full $\times$ no-SWT.no-2SH & $0.0045$ & $6.0\times 10^{-5}$ & $-0.021\pm 0.015$ & $-0.57$ \\
		Full $\times$ no-DSW.no-SWT & $0.005$ & $0.003$ & $-0.021\pm 0.019$ & $-0.41$ \\
		\hline
		\multicolumn{5}{c}{\textit{Stop rejecting $H_0$}}\\
		\hline
		Full $\times$ no-SHF.no-DSW & $0.0056$ & $0.03$ & $-0.014\pm 0.018$ & $-0.29$ \\
		Full $\times$ no-SHF.no-SWT & $0.0062$ & $0.038$ & $-0.0097\pm 0.013$ & $-0.28$ \\
		Full $\times$ no-DSW & $0.0071$ & $0.1$ & $-0.011\pm 0.019$ & $-0.22$ \\
		Full $\times$ no-SWP.no-2SH & $0.0083$ & $0.27$ & $-0.0089\pm 0.022$ & $-0.15$ \\
		Full $\times$ no-SHF & $0.01$ & $0.27$ & $-0.0045\pm 0.011$ & $-0.15$ \\
		Full $\times$ no-SWT & $0.012$ & $0.37$ & $~~0.0035\pm 0.0099$ & $0.12$ \\
		Full $\times$ no-SHF.no-SWP & $0.017$ & $0.6$ & $~~0.0033\pm 0.015$ & $0.071$ \\
		Full $\times$ no-SWP.no-SWT & $0.025$ & $0.65$ & $~~0.003\pm 0.015$ & $0.06$ \\
		Full $\times$ no-SWP & $0.05$ & $0.79$ & $~~0.0016\pm 0.012$ & $0.036$ \\
		\hline& 
	\end{tabular}
\label{tab:exp1res}
\end{table}

\begin{figure}[ht]
	\centering
	\includegraphics[width = \linewidth]{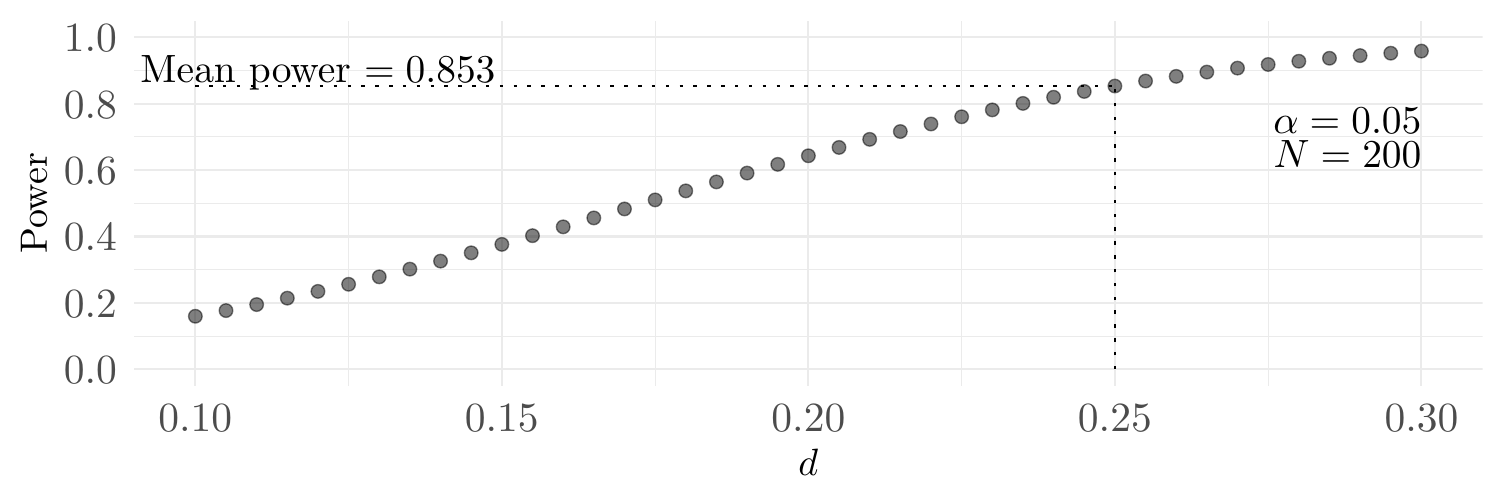}
	%\resizebox{.9\linewidth}{!}{\input{FinalPower.tex}}
	\caption{Power curve for the follow-up experiment, obtained by setting $N = 200$ and iterating over effect sizes to calculate the corresponding mean power of the experiment to detect different values of $d$.}
	\label{fig:power}
\end{figure}

To further explore these initial insights a follow-up experiment was performed. The objective of this second analysis was to obtain a more precise estimate of the effect (or lack thereof) of removing these three neighbourhood functions, individually as well as in pairs,  from the pool of movements available to the algorithm. Further, we also added the variant generated by simultaneously removing these three functions, which was not present in the first experiment. This means that this follow-up experiment involved 7 variants being compared against the \textit{Full} algorithm. The full $200$-instance test set was employed for this follow-up experiment, which provides a mean power of $\pi \approxeq 0.85$ to detect differences as small as $d^* = 0.25$. This was calculated by deriving the $d^*\times\pi^*$ curve shown in Figure \ref{fig:power}, using the same approach described in Section \ref{sec:ninstances}.

\clearpage

\begin{figure}[ht]
	\centering
	\includegraphics[width = \linewidth]{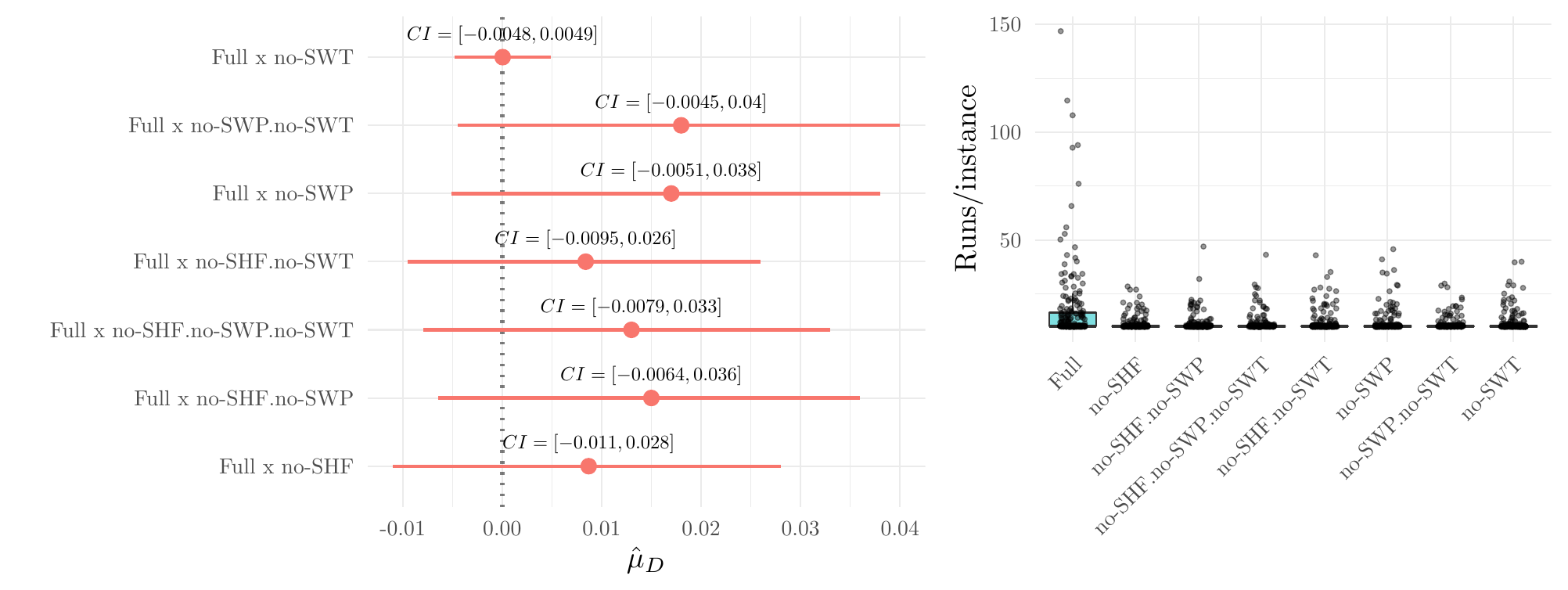}
	%\resizebox{\linewidth}{!}{\input{Final_CI_SS.tex}}
	\caption{Joint $95\%$ confidence intervals and sample sizes in the follow-up experiment. Notice that no comparison yielded statistically significant results, and that the narrow confidence intervals suggest that even if the neighbourhoods tested have some effect on the algorithm performance it is likely to be of minor consequence.}
	\label{fig:final_CI_SS}
\end{figure}

Table \ref{tab:exp1res2} and Figure \ref{fig:final_CI_SS} summarise the results obtained for the follow-up experiment. Notice that even with the increased sensitivity of the tests to the mean of percent differences for the variants that had the SWP, SHF and SWT neighbourhood functions removed from the pool of possible movements the tests failed to detect any statistically significant differences. This result, coupled with the narrow confidence intervals, indicates that any effect that these neighbourhood functions may be having on the performance of the algorithm are probably very small, if they exist at all. Based on these results, algorithm designers and developers could expand this exploration even further by, for instance, (i) focussing on the structures that are being explored by the remaining neighbourhood functions, so as to gain better understanding of the problem's landscape, or (ii) perform further exploration into instance subsets, to investigate whether the systematic lack of effect observed in this experiment also occurs, e.g., when conditioning the results on problem size. These further explorations are, however, outside the scope of the current paper.

\begin{table}[hb]
\centering\footnotesize
\caption{Summary of the 7 follow-up comparisons against the \textit{Full} algorithm. None of the results was statistically significant at the joint $95\%$ confidence level.}
\begin{tabular}{lllll}
 \hline
		Comparison & $\alpha_r^\prime$ & p-value & Confidence interval & $\widehat{d}~~$ \\
	 \hline
	 Full $\times$ no-SWP.no-SWT & $0.0071$ & $0.032$ & $0.018\pm 0.022$ & $0.15$ \\
	 Full $\times$ no-SWP & $0.0083$ & $0.043$ & $0.017\pm 0.022$ & $0.14$ \\
	 Full $\times$ no-SHF.no-SWP & $0.01$ & $0.072$ & $0.015\pm 0.021$ & $0.13$ \\
	 Full $\times$ no-SHF.no-SWP.no-SWT & $0.012$ & $0.12$ & $0.013\pm 0.021$ & $0.11$ \\
	 Full $\times$ no-SHF & $0.017$ & $0.28$ & $0.0087\pm 0.019$ & $0.077$ \\
	 Full $\times$ no-SHF.no-SWT & $0.025$ & $0.29$ & $0.0084\pm 0.018$ & $0.075$ \\
	 Full $\times$ no-SWT & $0.05$ & $0.98$ & $5.3\times 10^{-5}\pm 0.0048$ & $0.0015$ \\
	 \hline
\end{tabular}
\label{tab:exp1res2}
\end{table}

\section{Conclusions}
\label{sec:conclusions}

In this work we expanded methodology first introduced in \cite{Campelo2019} to enable sample size calculations for comparative experiments involving an arbitrary number of algorithms for the solution of a problem class of interest. The two aspects of sample size estimation for the comparative performance experiments, namely the number of instances and the amount of runs allocated to each algorithm on each instance, were addressed in a statistically principled manner. 

The number of instances is determined based on the desired sensitivity (in terms of statistical power) for the detection of effects larger than a predefined \textit{minimally relevant effect size} (MRES), under a given family-wise confidence level. Holm's step-down procedure was employed to control the rate of false positives while enabling the design of experiments at desired levels of power for the best, worst, median or mean case. The proposed methodology is based on the assumptions that underlie the t-tests, but the results can be easily generalised to the most common non-parametric tests (e.g., Wilcoxon signed-ranks test) based on asymptotic relative efficiency, as discussed previously in \cite{Campelo2019}.

The proposed iterative sampling method used to determine the number of runs of each algorithm on each instance is based on the interpretation of the standard error of estimation as a measure of precision, and generating samples aimed at minimising the total amount of runs necessary to obtain estimates with standard errors below predefined thresholds. Optimal sample size ratios were derived for simple and percent differences between the mean of all pairs of algorithms on any given instance, and a sampling heuristic was proposed that enables generalisation for other statistics that may be of eventual interest to researchers, such as the variance, algorithm rankings etc.

An example of application was provided in Section \ref{sec:application}, using the publicly available \texttt{R} package \texttt{CAISEr} \cite{caiser}, which provides an easy to use implementation of the proposed methodology. The example focussed on the comparison of a state-of-the-art heuristic for the solution of a class of scheduling problems against 21 algorithm variants generated by suppressing different neighbourhood functions used to generate candidate solutions. The preliminary results indicated not only the most relevant neighbourhood function but also the apparent lack of effect of three of the six ones commonly employed. The first part of the experiment illustrated the ability of the proposed method to sample the algorithms so as to control the standard errors at each instance under the desired accuracy threshold. It also showcased the compliance between the MRES used for designing the experiment, and the effect sizes that the hypothesis tests were able to detect as statistically significant.

The results obtained in the first part of the experiment were used to design a follow-up investigation, in which the effects of the three neighbourhood structures flagged in the initial part as not statistically significant were further investigated. This follow-up experiment used the full available instance set, and was used to showcase the use of the proposed methodology in a fixed sample size situation. The results obtained not only corroborated those obtained in the first part, but further reinforced the tentative conclusion that the effect of using the three neighbourhood functions as part of the algorithm must be minor at best, which suggests a few possibilities of exploration for algorithm designers and developers.

% Adicionar intervalos de confiança aos valores de d reportados nas tabelas 1 e 2 (maybe)

\subsection{Limitations and Possibilities}
As in our previous work upon which the present paper was based, it is important to reinforce at that the proposed methodology is not the definitive way to compare algorithms. For instance, convergence analysis, algorithm reliability, and performance analysis conditional on problem characteristics are research questions that require different methodologies. The proposed methodology is simply an additional tool in the research arsenal, one that can provide answers to several common questions in the experimental comparison of algorithms.

One of the most straightforward extensions of the work presented in this paper is the incorporation of sequential analysis approaches \cite{Botella2006,Bartroff2013} to enable further reductions in the required number of instances, which is possible in cases where the actual effect sizes are substantially larger than the MRES defined in the experimental design. Extending the sample size calculations performed here to Bayesian alternatives \cite{Calvo2019} is also relatively straightforward, and may allow the aggregation of existing knowledge  into the comparison of algorithms, in the form of prior distributions. Bayesian methods may also allow for an easier incorporation of sequential analysis into the sample size calculation methodologies - since the incremental aggregation of observations in Bayesian inference does not require further significance corrections  \cite{Kruschke2010} - and may even enable the incorporation of existing published evidence, which would be another step towards the development of meta-analytical tools for algorithmic research.

\bibliographystyle{spmpsci}      % mathematics and physical sciences
%\bibliography{02_bibliography}   % name your BibTeX data base

\begin{thebibliography}{10}
\providecommand{\url}[1]{{#1}}
\providecommand{\urlprefix}{URL }
\expandafter\ifx\csname urlstyle\endcsname\relax
  \providecommand{\doi}[1]{DOI~\discretionary{}{}{}#1}\else
  \providecommand{\doi}{DOI~\discretionary{}{}{}\begingroup
  \urlstyle{rm}\Url}\fi

\bibitem{Amo2012}
del Amo, I.G., Pelta, D.A., Gonz{\'{a}}lez, J.R., Masegosa, A.D.: An algorithm
  comparison for dynamic optimization problems.
\newblock Applied Soft Computing \textbf{12}(10), 3176--3192 (2012)

\bibitem{Barr1995}
Barr, R.S., Golden, B.L., Kelly, J.P., Resende, M.G.C., Stewart, W.R.:
  Designing and reporting on computational experiments with heuristic methods.
\newblock Journal of Heuristics \textbf{1}(1), 9--32 (1995)

\bibitem{Bartroff2013}
Bartroff, J., Lai, T., Shih, M.C.: Sequential Experimentation in Clinical
  Trials: Design and Analysis.
\newblock Springer (2013)

\bibitem{BartzBeielstein2005}
Bartz-Beielstein, T.: {New Experimentalism Applied to Evolutionary
  Computation}.
\newblock Ph.D. thesis, Universit\"at Dortmund, Germany (2005)

\bibitem{Bartz-Beielstein2006}
Bartz-Beielstein, T.: Experimental Research in Evolutionary Computation.
\newblock Springer (2006)

\bibitem{BartzBeielstein2015}
Bartz-Beielstein, T.: How to create generalizable results.
\newblock In: J.~Kacprzyk, W.~Pedrycz (eds.) Handbook of Computational
  Intelligence. Springer (2015)

\bibitem{Bartz-Beielstein2010}
Bartz-Beielstein, T., Chiarandini, M., Paquete, L., Preuss, M.: Experimental
  Methods for the Analysis of Optimization Algorithms.
\newblock Springer (2010)

\bibitem{Benavoli2014}
Benavoli, A., Corani, G., Mangili, F., Zaffalon, M., Ruggeri, F.: A bayesian
  wilcoxon signed-rank test based on the dirichlet process.
\newblock In: 30th International conference on machine learning, pp. 1026--1034
  (2014)

\bibitem{Birattari2004}
Birattari, M.: On the estimation of the expected performance of a metaheuristic
  on a class of instances: how many instances, how many runs?
\newblock Tech. Rep. IRIDIA/2004-001, Université Libre de Bruxelles, Belgium
  (2004)

\bibitem{Birattari2009}
Birattari, M.: Tuning Metaheuristics -- A Machine Learning Perspective.
\newblock Springer Berlin Heidelberg (2009)

\bibitem{Birattari2007}
Birattari, M., Dorigo, M.: {How to assess and report the performance of a
  stochastic algorithm on a benchmark problem: Mean or best result on a number
  of runs?}
\newblock Optimization Letters \textbf{1}, 309--311 (2007)

\bibitem{Botella2006}
Botella, J., Xim{\'{e}}nez, C., Revuelta, J., Suero, M.: Optimization of sample
  size in controlled experiments: The {CLAST} rule.
\newblock Behavior Research Methods \textbf{38}(1), 65--76 (2006)

\bibitem{Calvo2019}
Calvo, B., Shir, O.M., Ceberio, J., Doerr, C., Wang, H., B\"{a}ck, T., Lozano,
  J.A.: Bayesian performance analysis for black-box optimization benchmarking.
\newblock In: Proceedings of the Genetic and Evolutionary Computation
  Conference, GECCO '19, pp. 1789--1797. ACM (2019)

\bibitem{caiser}
Campelo, F.: CAISEr: Comparison of Algorithms with Iterative Sample Size
  Estimation (2019).
\newblock \urlprefix\url{https://CRAN.R-project.org/package=CAISEr}.
\newblock Package version 1.0.13

\bibitem{Campelo2019}
Campelo, F., Takahashi, F.: Sample size estimation for power and accuracy in
  the experimental comparison of algorithms.
\newblock Journal of Heuristics \textbf{25}(2), 305--338 (2019)

\bibitem{Carrano2011}
Carrano, E.G., Wanner, E.F., Takahashi, R.H.C.: A multicriteria statistical
  based comparison methodology for evaluating evolutionary algorithms.
\newblock {IEEE} Transactions on Evolutionary Computation \textbf{15}(6),
  848--870 (2011)

\bibitem{Chimani2010}
Chimani, M., Klein, K.: Algorithm engineering: Concepts and practice.
\newblock In: Experimental Methods for the Analysis of Optimization Algorithms,
  pp. 131--158. Springer Berlin Heidelberg (2010)

\bibitem{CoffinSaltzman2000}
Coffin, M., Saltzman, M.J.: Statistical analysis of computational tests of
  algorithms and heuristics.
\newblock {INFORMS} Journal on Computing \textbf{12}(1), 24--44 (2000)

\bibitem{Czarn2004}
Czarn, A., MacNish, C., Vijayan, K., Turlach, B.: Statistical exploratory
  analysis of genetic algorithms: the importance of interaction.
\newblock In: Proceedings of the 2004 {IEEE} Congress on Evolutionary
  Computation. Institute of Electrical {\&} Electronics Engineers ({IEEE})
  (2004)

\bibitem{Demsar2006}
Dem\v{s}ar, J.: Statistical comparisons of classifiers over multiple data sets.
\newblock Journal of Machine Learning Research \textbf{7}, 1--30 (2006)

\bibitem{Derrac2014}
Derrac, J., Garc{\'{i}}a, S., Hui, S., Suganthan, P.N., Herrera, F.: Analyzing
  convergence performance of evolutionary algorithms: A statistical approach.
\newblock Information Sciences \textbf{289}, 41--58 (2014)

\bibitem{Derrac2011}
Derrac, J., Garc{\'{\i}}a, S., Molina, D., Herrera, F.: A practical tutorial on
  the use of nonparametric statistical tests as a methodology for comparing
  evolutionary and swarm intelligence algorithms.
\newblock Swarm and Evolutionary Computation \textbf{1}(1), 3--18 (2011)

\bibitem{Dunn1961}
Dunn, O.J.: Multiple comparisons among means.
\newblock Journal of the American Statistical Association \textbf{56}(293),
  52--64 (1961)

\bibitem{Eiben2002}
Eiben, A., Jelasity, M.: A critical note on experimental research methodology
  in {EC}.
\newblock In: Proceedings of the 2002 {IEEE}Congress on Evolutionary
  Computation. Institute of Electrical {\&} Electronics Engineers ({IEEE})
  (2002)

\bibitem{Ellis2010}
Ellis, P.D.: The Essential Guide to Effect Sizes, 1st edn.
\newblock Cambridge University Press (2010)

\bibitem{Fieller1954}
Fieller, E.C.: Some problems in interval estimation.
\newblock Journal of the Royal Statistical Society. Series B (Methodological)
  \textbf{16}(2), 175--185 (1954)

\bibitem{Franz2007}
Franz, V.: Ratios: A short guide to confidence limits and proper use (2007).
\newblock \url{https://arxiv.org/pdf/0710.2024v1.pdf}

\bibitem{Garcia2009}
Garc{\'{\i}}a, S., Fern{\'{a}}ndez, A., Luengo, J., Herrera, F.: A study of
  statistical techniques and performance measures for genetics-based machine
  learning: accuracy and interpretability.
\newblock Soft Computing \textbf{13}(10), 959--977 (2009)

\bibitem{Garcia2010}
Garc{\'{\i}}a, S., Fern{\'{a}}ndez, A., Luengo, J., Herrera, F.: Advanced
  nonparametric tests for multiple comparisons in the design of experiments in
  computational intelligence and data mining: Experimental analysis of power.
\newblock Information Sciences \textbf{180}(10), 2044--2064 (2010)

\bibitem{Garcia2008}
Garc{\'{\i}}a, S., Molina, D., Lozano, M., Herrera, F.: A study on the use of
  non-parametric tests for analyzing the evolutionary algorithms' behaviour: a
  case study on the {CEC}'2005 {S}pecial session on real parameter
  optimization.
\newblock Journal of Heuristics \textbf{15}(6), 617--644 (2008)

\bibitem{Gelman2006}
Gelman, A., Hill, J.: Data analysis using regression and
  multilevel/hierarchical models.
\newblock Cambridge university press (2006)

\bibitem{Graham1979}
Graham, R.L., Lawler, E.L., Lenstra, J.K., {Rinnooy Kan}, A.H.G.: Optimization
  and approximation in deterministic sequencing and scheduling: a survey.
\newblock Annals of Discrete Mathematics \textbf{5}, 287--326 (1979)

\bibitem{COCO2016}
Hansen, N., T\v{u}sar, T., Mersmann, O., Auger, A., Brockoff, D.: {COCO}: The
  experimental procedure (2016).
\newblock \urlprefix\url{https://arxiv.org/abs/1603.08776}

\bibitem{Holm1979}
Holm, S.: A simple sequentially rejective multiple test procedure.
\newblock Scandinavian journal of statistics \textbf{6}(2), 65--70 (1979)

\bibitem{Hooker1994}
Hooker, J.N.: Needed: An empirical science of algorithms.
\newblock Operations Research \textbf{42}(2), 201--212 (1994)

\bibitem{Hooker1996}
Hooker, J.N.: Testing heuristics: We have it all wrong.
\newblock Journal of Heuristics \textbf{1}(1), 33--42 (1996)

\bibitem{Hurlbert1984}
Hurlbert, S.H.: Pseudoreplication and the design of ecological field
  experiments.
\newblock Ecological Monographs \textbf{54}(2), 187--211 (1984)

\bibitem{Jain1991}
Jain, R.K.: The Art of Computer Systems Performance Analysis.
\newblock John Wiley and Sons Ltd (1991)

\bibitem{Johnson2002}
Johnson, D.: A theoretician's guide to the experimental analysis of algorithms.
\newblock In: M.~Goldwasser, D.~Johnson, C.~McGeoch (eds.) Data Structures,
  Near Neighbor Searches, and Methodology: Fifth and Sixth DIMACS
  Implementation Challenges, \emph{DIMACS Series in Discrete Mathematics and
  Theoretical Computer Science}, vol.~59, pp. 215--250. American Mathematical
  Society (2002)

\bibitem{Krohling2015}
Krohling, R.A., Lourenzutti, R., Campos, M.: Ranking and comparing evolutionary
  algorithms with hellinger-{TOPSIS}.
\newblock Applied Soft Computing \textbf{37}, 217--226 (2015)

\bibitem{Kruschke2010}
Kruschke, J.K.: Doing Bayesian Data Analysis: A Tutorial with R and BUGS, 1st
  edn.
\newblock Academic Press, Inc. (2010)

\bibitem{Lawler1993}
Lawler, E.L., Lenstra, J.K., {Rinnooy Kan}, A.H., Shmoys, D.B.: Sequencing and
  scheduling: Algorithms and complexity.
\newblock In: Handbooks in Operations Research and Management Science, vol.~4,
  chap.~9, pp. 445--522. Elsevier (1993)

\bibitem{Lazic2010}
Lazic, S.E.: The problem of pseudoreplication in neuroscientific studies: is it
  affecting your analysis?
\newblock {BMC} Neuroscience \textbf{11}(5), 397--407 (2010)

\bibitem{Lenth2001}
Lenth, R.V.: Some practical guidelines for effective sample size determination.
\newblock The American Statistician \textbf{55}(3), 187--193 (2001)

\bibitem{Maravilha2019}
Maravilha, A.L., Pereira, L.M., Campelo, F.: Statistical characterization of
  neighborhood structures for the unrelated parallel machine problem with
  sequence-dependent setup times.
\newblock In preparation

\bibitem{Mathews2010}
Mathews, P.: Sample Size Calculations: Practical Methods for Engineers and
  Scientists, 1st edn.
\newblock Matthews Malnar \& Bailey Inc. (2010)

\bibitem{McGeoch1996}
McGeoch, C.C.: Feature article{\textemdash}toward an experimental method for
  algorithm simulation.
\newblock {INFORMS} Journal on Computing \textbf{8}(1), 1--15 (1996)

\bibitem{Millar2004}
Millar, R., Anderson, M.: Remedies for pseudoreplication.
\newblock Fisheries Research \textbf{70}, 397--407 (2004)

\bibitem{Montgomery2013b}
Montgomery, D.C.: Design and Analysis of Experiments, 8th edn.
\newblock John Wiley \& Sons (2013)

\bibitem{Montgomery2013}
Montgomery, D.C., Runger, G.C.: Applied Statistics and Probability for
  Engineers, 6th edn.
\newblock Wiley (2013)

\bibitem{Pereira2019}
Pereira, L.M.: {An\'alise de Estruturas de Vizinhan\c{c}a para o Problema de
  Sequenciamento de M\'aquinas Paralelas N\~ao Relacionadas com Tempos de
  Prepara\c{c}\~ao }.
\newblock Master's thesis, Universidade Federal de Minas Gerais, Belo
  Horizonte, Brazil (2019).
\newblock \urlprefix\url{https://ppgee.ufmg.br/defesas/1615M.PDF}.
\newblock In Portuguese

\bibitem{R}
{R Core Team}: R: A Language and Environment for Statistical Computing.
\newblock R Foundation for Statistical Computing, Vienna, Austria (2017).
\newblock \urlprefix\url{https://www.R-project.org/}

\bibitem{Ridge2007}
Ridge, E.: {Design of Experiments for the Tuning of Optimisation Algorithms}.
\newblock Ph.D. thesis, The University of York, UK (2007)

\bibitem{Santos2016}
Santos, H.G., Toffolo, T.A., Silva, C.L., Berghe, G.V.: Analysis of stochastic
  local search methods for the unrelated parallel machine scheduling problem.
\newblock International Transactions in Operational Research \textbf{00} (2016)

\bibitem{Shaffer1995}
Shaffer, J.P.: Multiple hypothesis testing.
\newblock Annual review of psychology \textbf{46}(1), 561--584 (1995)

\bibitem{Sheskin2011}
Sheskin, D.J.: Handbook of Parametric and Nonparametric Statistical Procedures.
\newblock Taylor \& Francis (2011)

\bibitem{Sorensen2018}
S\"{o}rensen, K., Sevaux, M., Glover, F.: A history of metaheuristics.
\newblock In: Handbook of Heuristics, pp. 1--18. Springer International
  Publishing (2018)

\bibitem{Vallada2011}
Vallada, E., Ruiz, R.: A genetic algorithm for the unrelated parallel machine
  scheduling problem with sequence dependent setup times.
\newblock European Journal of Operational Research \textbf{211}(3), 612--622
  (2011)

\bibitem{Yuan2004}
Yuan, B., Gallagher, M.: {Statistical racing techniques for improved empirical
  evaluation of evolutionary algorithms}.
\newblock Parallel Problem Solving From Nature - {PPSN VIII} \textbf{3242},
  172--181 (2004)

\bibitem{YuanGallagher2009}
Yuan, B., Gallagher, M.: An improved small-sample statistical test for
  comparing the success rates of evolutionary algorithms.
\newblock In: Proceedings of the 11th Annual conference on Genetic and
  evolutionary computation - {GECCO}09. Association for Computing Machinery
  ({ACM}) (2009)

\end{thebibliography}

\end{document}